\documentclass[final,1p,authoryear]{elsarticle}

\usepackage{amssymb}
\usepackage{amsmath}

\usepackage{gensymb}

\usepackage{lineno}

\journal{Planetary and Space Science}

\begin{document}

\begin{frontmatter}



\title{Dynamical regimes of two eccentric and mutually inclined giant planets} 


\author{Tabaré Gallardo} 
\ead{tabare.gallardo@fcien.edu.uy}
\author{Alfredo Suescun} 
\ead{alfredosuescun@gmail.com}

\affiliation{organization={Facultad de Ciencias, Udelar},
            addressline={Igua 4225}, 
            postcode={11400}, 
            state={Montevideo},
            country={Uruguay}}

\begin{abstract}
We consider a basic planetary system composed by a Sun like star, a Jupiter-like planet an a Neptune-like planet in a wide range of orbital configurations not limited to the hierarchical case. We present atlases of resonances showing the domains of $\sim 1300$ mutual mean-motion resonances (MMRs)  and their link to chaotic and regular dynamics.  Following a semi-analytical method for the study of the secular dynamics we found two regimes for equilibrium configurations: one for low mutual inclinations were equilibrium is related to oscillations of $\Delta\varpi$ around $0\degree$ or $180\degree$, and other for high mutual inclinations where the equilibrium is given by defined values of the $\omega_i$ equal to integer multiples of $90\degree$.
By numerical integration of the full equations of motion we calculate the fundamental frequencies of the systems in their diverse configurations and study their dependence with the orbital elements. According to the analysis of the fundamental frequencies we found two dynamical regimes depending on the initial mutual inclination and the limit between the two regimes occurs at some critical inclination $30\degree \lesssim i_c  \lesssim 40\degree$ defined by the occurrence of the secular resonance $g_1=g_2$. For $i<i_c$  the dynamics is analogue a the classic secular model for low $(e,i)$ with well defined three fundamental frequencies and free and forced modes, conserving quasi constant the mutual inclination. For $i>i_c$ the dynamics is completely different with increasing changes in mutual inclination and emerging combinations of the fundamental frequencies and, depending on the case, dominated by the secular resonance or the vZLK mechanism.

\end{abstract}

\begin{graphicalabstract}
 \includegraphics[width=1\linewidth]{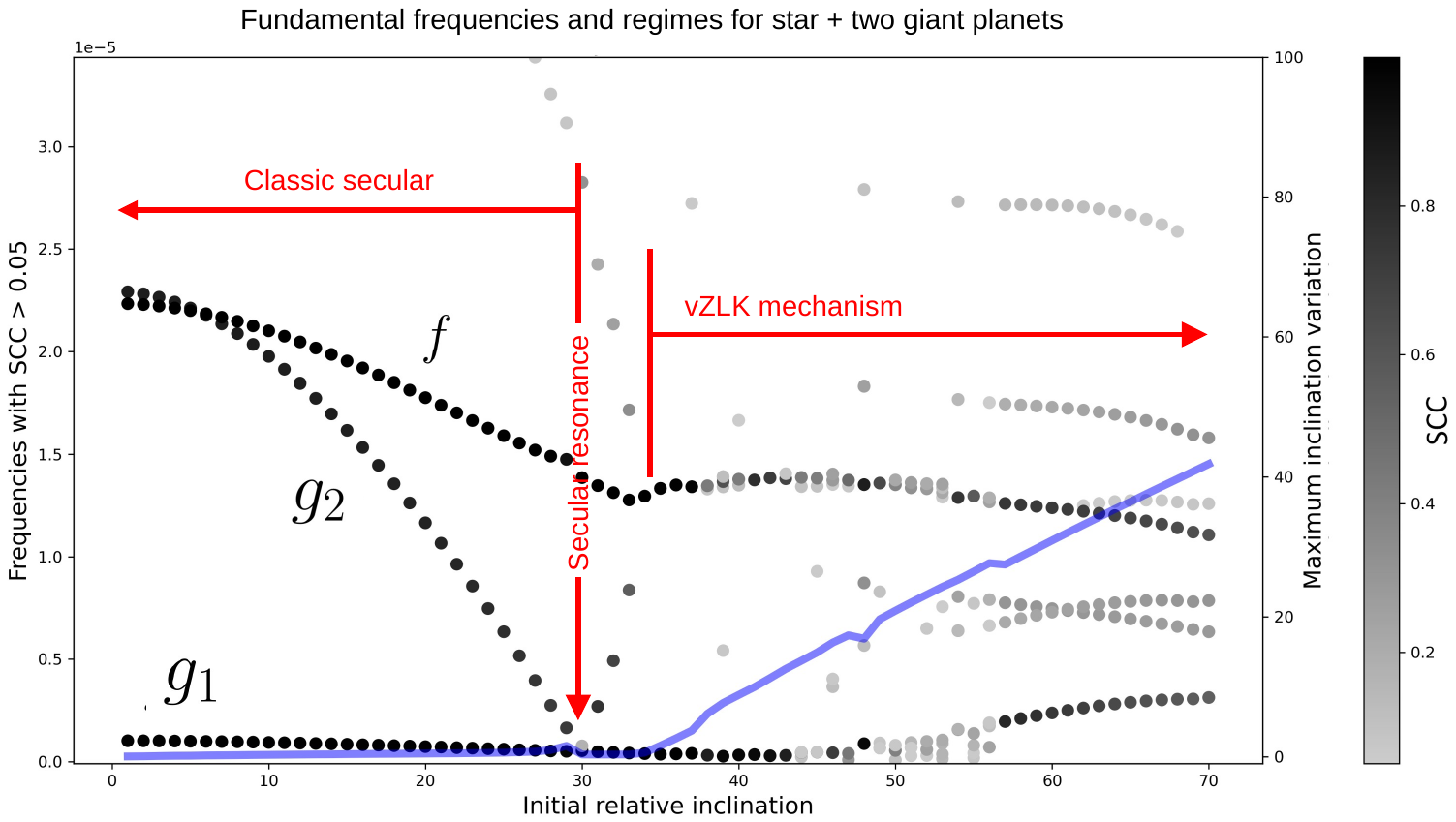}
 
\end{graphicalabstract}

%
%
%
%
%
%
%
%
%
%


\end{frontmatter}




\section{Introduction}

The long-term dynamics of planetary systems is characterized by secular theories that, with a minimum of variables, manage to describe the essential aspects of the orbital evolution. 
 Since the fundamental work of Euler, Laplace and Lagrange other secular theories for planetary systems with low $(e,i)$-orbits were developed  more recently \citep{1986A&A...157...59L}.
For arbitrary $(e,i)$-orbits the mutual disturbing function between the planets becomes unmanageable and the equations impossible to solve unless some strong simplification of the problem is made.
Taking advantage of the growing computation power another approach to compute the long term dynamics was done by means of synthetic theories \citep{CARPINO1987} which are based on filtered numerical integration of the equations of motion and give very good description of the secular evolution of planetary systems but they do not provide an understanding of the dynamical mechanism that drive the system's evolution.

If we want to analyze the most general problem including arbitrary planetary orbits, the simplest case is that of a star with two planets.
This case must be understood well before trying to delve into more complex systems.
The two planet planar case is the most simple one because can be reduced to only one degree of freedom, being $\Delta\varpi$, the difference between the  longitude of the pericenters of the planetary orbits, the only relevant angular variable. Both planetary eccentricities are related by the conservation of the system's angular momentum.
The only problem is how to handle analytically the equations which we pretend to be valid for a wide range of orbital eccentricities, considering the great diversity observed in the exoplanets. For widely separated planetary orbits, called hierarchical case, it is possible to obtain quite good approximations for the mutual disturbing function, based on low order expansions in terms of $a_1/a_2$.
The secular dynamics for the planar hierarchical case was analytically studied for example by \cite{2003ApJ...592.1201L} using quadrupole and octupole models. They found the conditions, in terms of the ratio between the planet's angular momenta, for the existence of equilibrium points around $\Delta\varpi=0\degree$ and $180\degree$, and possible trajectories in the space $(e_i,\Delta\varpi)$. In particular they studied the system HD12661 which is the first system where $\Delta\varpi$ was found to oscillate.

Almost simultaneously \cite{2004Icar..168..237M}
studied the same problem but with a very different approach: computing numerically, not analytically, the disturbing function, or more properly, the Hamiltonian of the system, allowing the study of high eccentricity orbits. Considering that extrasolar systems frequently exhibit large orbital eccentricities the numerical calculation of the secular disturbing function becomes essential to obtain reliable results, avoiding the limitations of the analytical series expansions of the disturbing functions.
They analyzed the system in the called \textit{representative plane} which shows the values of the eccentricities of the planets for the specific instant when  $\Delta\varpi$ is $0\degree$ or $180\degree$. They studied equilibrium points finding the families called Mode I (oscillations of $\Delta\varpi$ around $0\degree$) and Mode II (oscillations of $\Delta\varpi$ around $180\degree$). It is worth noting  they state the difference between apsidal libration, that means oscillations of $\Delta\varpi$ and secular resonances which are librations surrounded by a separatrix with a bifurcation point, situation that in the planar case  happens mainly at high eccentricities.

Using the same semi-analytical approach 
\citet{2006Icar..181..555M} studied the 
more complex three-dimensional (3D) case. This work represents a milestone in the study of the planetary 3D dynamics. Using a convenient set of variables they showed it is possible to reduce the system to a two degrees of freedom and they state the most important dynamical properties of  two-planets systems. For example, they demonstrate that the general properties do not depend on individual values of semimajor axes and masses but in their ratios. They report regions of oscillations of $\Delta\varpi$ around $0\degree$ and $180\degree$, regimes of oscillations of the $\omega_i$ around $\pm 90\degree$ following a vZLK (von Zeipel - Lidov - Kozai) regime and chaotic regimes in the transition regions.
\citet{2009MNRAS.395.1777M} followed a similar approach but focused  on the equilibrium points, that means points of the space of orbital elements where the system remains unchanged over time, and in obtaining families of periodic solutions. In their exhaustive study they made a complete survey and
found new equilibrium solutions. The 3D secular dynamics of 2 planets following analytical classic expansions extended to high order was done for example by \citet{2007thesislibert}, \cite{2008CeMDA.100..209L}, \citet{2009A&A...493..677L}. Although these developments do not allow studies of very high eccentric systems, they are useful for analyzing the dynamics of many systems and provide an extension of the classic Lagrange-Laplace theory. It is interesting that they found analytical expressions for the fundamental frequencies, matter strongly related to the focus of our work.

The 3D hierarchical case using the more simple quadrupole and octupole models valid for small values of $a_1/a_2$ were studied for example by \cite{Naoz2013} and
\cite{Hansen2020}. In this last work the authors made an exhaustive study of equilibrium points that in spite of being valid for the hierarchical case it provides a very good description of the variety of the stationary points that the 3D planetary case can offer.  Employing multipole expansions, \cite{2023CeMDA.135...22M} investigated the hierarchical three-body problem and identified two general dynamical regimes: a 'planar-like' regime exhibiting dynamics similar to the planar case, and a 'Lidov-Kozai' regime, the occurrence of which depends on the mutual orbital inclination.

Another very different line of work is the study of the stability using a broad spectrum of numerical techniques.
The stability of planetary systems 
based on AMD stability criteria was studied by \cite{Laskar2017}, \cite{Petit2018}, \cite{He2020} and
\cite{Turrini2020}. Also using several dynamical criteria as in
\cite{2011A&A...526A..98F}, \cite{2019A&A...626A..74V},  \cite{Tamayo2020}, \cite{Gajdos2022}, \cite{Bhaskar2024a}, \citet{bhaskar2024} and \citet{Volk2024}.
 In particular \cite{2011A&A...526A..98F} and \cite{2019A&A...626A..74V}
found that instabilities arise when the mutual planetary inclination is large enough, and in general they establish this limit around $30\degree - 40\degree$. 
It is also worth mentioning that  there are also some studies about circumbinary planets like \cite{Lei2024} where analytical and numerical approaches are presented identifying secular resonances through a methodology based on the $||\Delta D||$ new dynamic indicator \citep{Daquin2023}. Finally, it is important to state that there are several studies of the restricted version of this problem, that means, when the small planet has no effect on the large planet. Even when some results of the restricted case can be analogue to the general case, the problem is different from the starting point.

In summary, there are currently a good number of studies that describe and explain the main aspects of the dynamics of two-planet systems, in particular for the hierarchical case. In this work we aim to provide complementary results from a different perspective and in particular in the range of semi-major axis ratio where the hierarchical models cannot be applied and with planetary mass ratio more close to the unity making mutual perturbations essential to the dynamics.
We will look for some general properties of the dynamics of eccentric and mutually inclined planets without the limitations given by analytical theories based on series expansions or by the restricted three body problem approximation.
To achieve this we will work with numerical integrations of the exact equations of motion and with two semi-analytical models, one for MMRs based on \cite{Gallardo2021} and the other   based in the numerical calculation of the secular Hamiltonian, following the ideas of \citet{2006Icar..181..555M}. We focus on a simple planetary system composed by a star with mass $m_0=1 M_{\odot}$ and two planets: one fixed  Jupiter-like ($m_J=0.001$) at $a_J=8$ au and one Neptune-like planet with $1<a_N<20$ au and nominal mass $m_N=m_J/10$, but we also explore the influence of the mass ratio $m_N/m_J$ in the dynamics.  Our study is founded in the numerical determination of the fundamental frequencies of the system, how they depend on the eccentricities and mutual inclination and in how they determine the dynamical evolution. To achieve this, first, in Section 2 we present a study of the resonance domains to clearly define the expected region for secular and resonant evolution. In Section 3 we develop our main study of the secular dynamics and in Section 4 we summarize the conclusions. We exclude from our study the more complex case of the secular evolution inside a MMR. It is known that the long term evolution inside a MMR produces large orbital changes for the planar case \citep{pons2024} but the complexity of the 3D case exceeds the intentions of this work.

\section{Regime dominated by mean motion resonances}

\begin{figure}[hbt!]
	\centering    \includegraphics[width=1\linewidth]{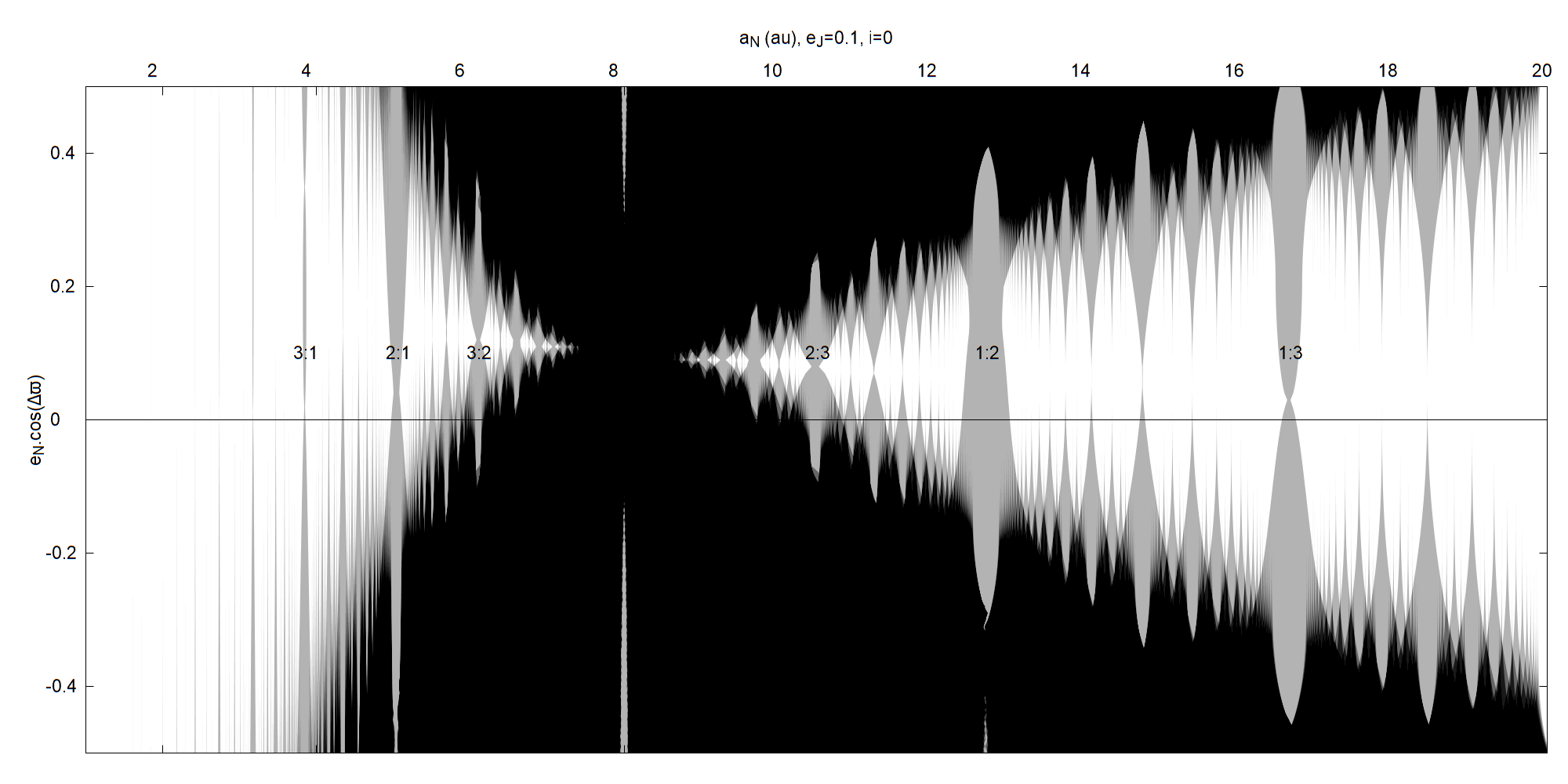}
	\caption{Atlas of MMRs $k_1/k_2$ with $k_i\leq 50$ in the plane ($a_N,e_N$) assuming a  Jupiter planet ($m_J=0.001$) at $a_J=8$ au, with $e_J=0.1$ and coplanar with the Neptune-like planet ($m_N=m_J/10$). Two extreme apsidal configurations are showed: $\Delta\varpi=0\degree$ at superior part and $\Delta\varpi=180\degree$ inferior. Grey regions correspond to stable resonances, black ones correspond to overlap of resonances. White regions are free from MMRs.}
	\label{atlasae}
\end{figure}

\begin{figure}[hbt!]
	\centering    \includegraphics[width=1\linewidth]{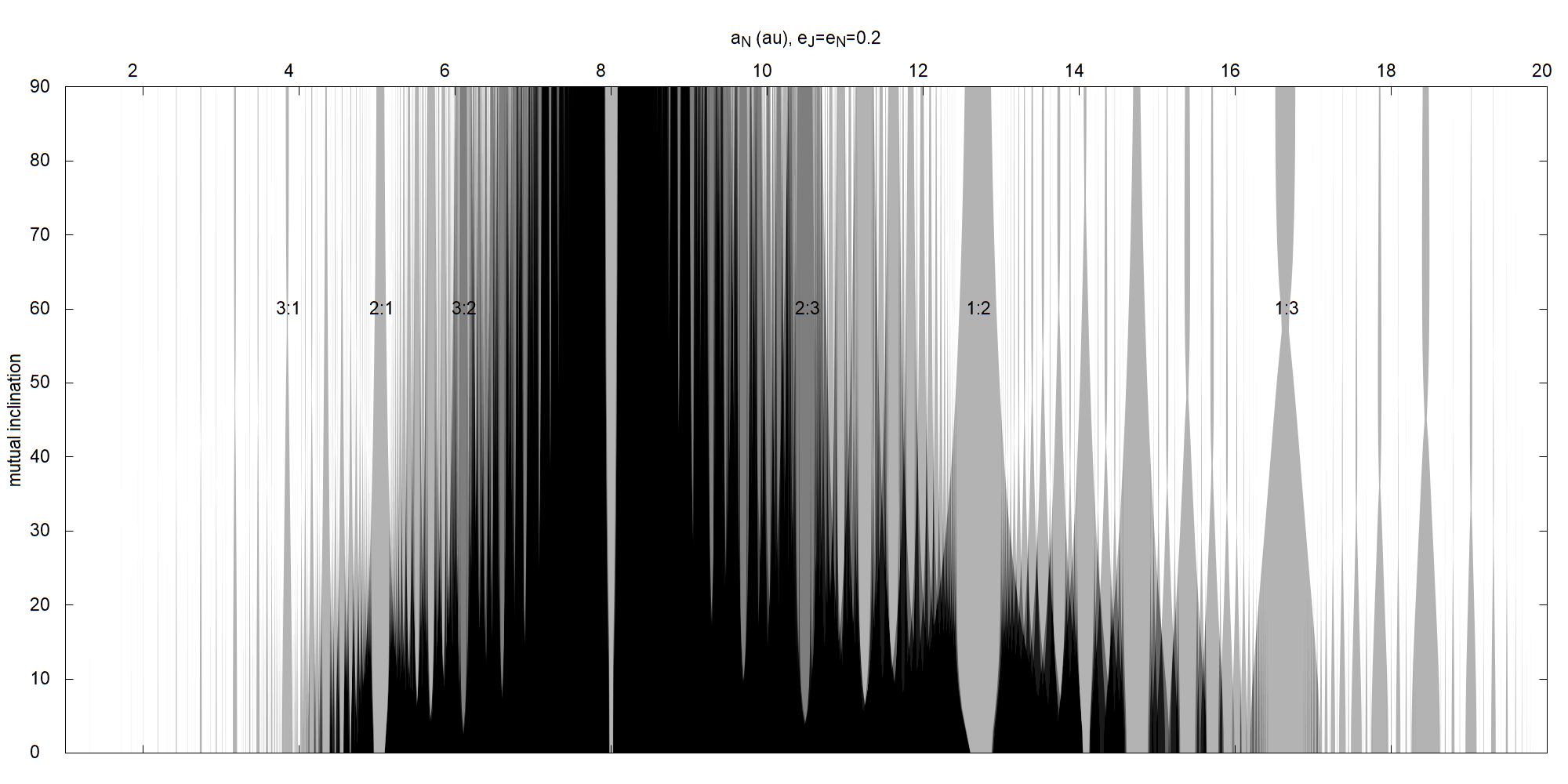}
	\caption{Same as Fig. \ref{atlasae} in the plane $(a_N,i)$, in this case  with $e_J=e_N=0.2$ and  $\Delta\varpi=180\degree$.  }
	\label{atlasai}
\end{figure}

In our study we will consider a star with mass  $m_0=1 M_{\odot}$, a Jupiter-like planet that in all cases have mass $m_J=0.001$ and semi-major axis $a_J=8$ au and a Neptune-like planet with $m_N=m_J/10$. Note that the nominal mass we adopted for our Neptune-like planet is approximately twice the actual mass of Neptune.
Using the model proposed by \cite{Gallardo2021} and the codes provided in the MMRs site\footnote{https://sites.google.com/view/mmresonances/} 
 we 
explored the domains of all MMRs between both planets verifying  $k_1n_1\simeq k_2n_2$ with the integers $k_i\leq 50$, where the $n_i$ are the mean motions,  in the space $(a_N,e_N,i)$ where $i$ is the mutual inclination and we present the results in Figs. \ref{atlasae} and \ref{atlasai}. A total of 1324 MMRs are showed in each figure. 
 The model is based on the numerical calculation of the resonant Hamiltonian so it works for arbitrary values of eccentricity and inclination. 
The domain of each resonance in the plane $(a_N,e_N)$ in Fig. \ref{atlasae} were calculated for coplanar orbits in two extreme configurations: $\Delta\varpi=0\degree$ in the upper part and $180\degree$ in the lower part of the plot.
The domain  in the plane $(a_N,i)$ in Fig. \ref{atlasai} were calculated for  $e_N=e_J=0.2$  and just for  $\Delta\varpi=180\degree$  which is the configuration where the resonances are strongest and widest. Each resonance domain, defined by the maximum libration $\Delta a_N$, is painted in gray so black regions correspond to resonance overlap which most probably generate chaotic behavior. White regions indicate absence of MMRs, so in principle pure secular dynamics could happen there. Note as some strong resonances dominate like 2:1, 3:2 and specially the exteriors 1:2 and 1:3   and others survive in the seas of black chaotic  regions like the eccentric case of 2:1 for $\Delta\varpi=180\degree$  or the  inclined coorbitals shown in Fig. \ref{atlasai}.

In the plane  $(a_N,e_N)$ of Fig. \ref{atlasae} in general the domains are wider for higher eccentricities as expected and
the structures are shifted respect to $e_N=0$ because of a forced mode generated by the non zero eccentricity $e_J$ \citep{2023CeMDA.135....3G}. 
In the plane  $(a_N,i)$ of Fig. \ref{atlasai} the domains are narrower for higher inclinations also as expected because in the end the resonance strength and width are related to the strength of the gravitational interactions between both planets which diminishes with $i$. Note that 1:1 resonance is isolated at least in its central region, so theoretically a Neptune-like planet coorbital with Jupiter could evolve there. Libration periods go from $10^2$ yrs for the strongest resonances to $10^4$ yrs for the weakest ones.
These figures show that approximately in the interval $ 4\lesssim a_N \lesssim 16$ au for $e_i\sim 0.2$
	and  $i\lesssim 30\degree$ overlap of resonances dominates.

The atlas of Fig. \ref{atlasae} was  generated with a Jupiter planet with $e_J=0.1$, when taking greater $e_J$ the atlas show wider resonances  shifted to greater values of $e_N\cos\Delta\varpi$. This suggest that for high eccentricity giant planets there is more room for secular evolutions for $\Delta\varpi=0\degree$ respect to $\Delta\varpi=180\degree$, considering the differences in the resonances overlap on both parts of the Fig. \ref{atlasae}. 
There is an interesting example of a planetary system  with an eccentric Jupiter planet at $a=8$ au studied by \cite{2022AJ....163..273E}. In their Fig. 2 it is possible to find several structures generated by MMRs with the giant planet HD 83443c.

\begin{figure}[hbt!]
	\centering    \includegraphics[width=1\linewidth]{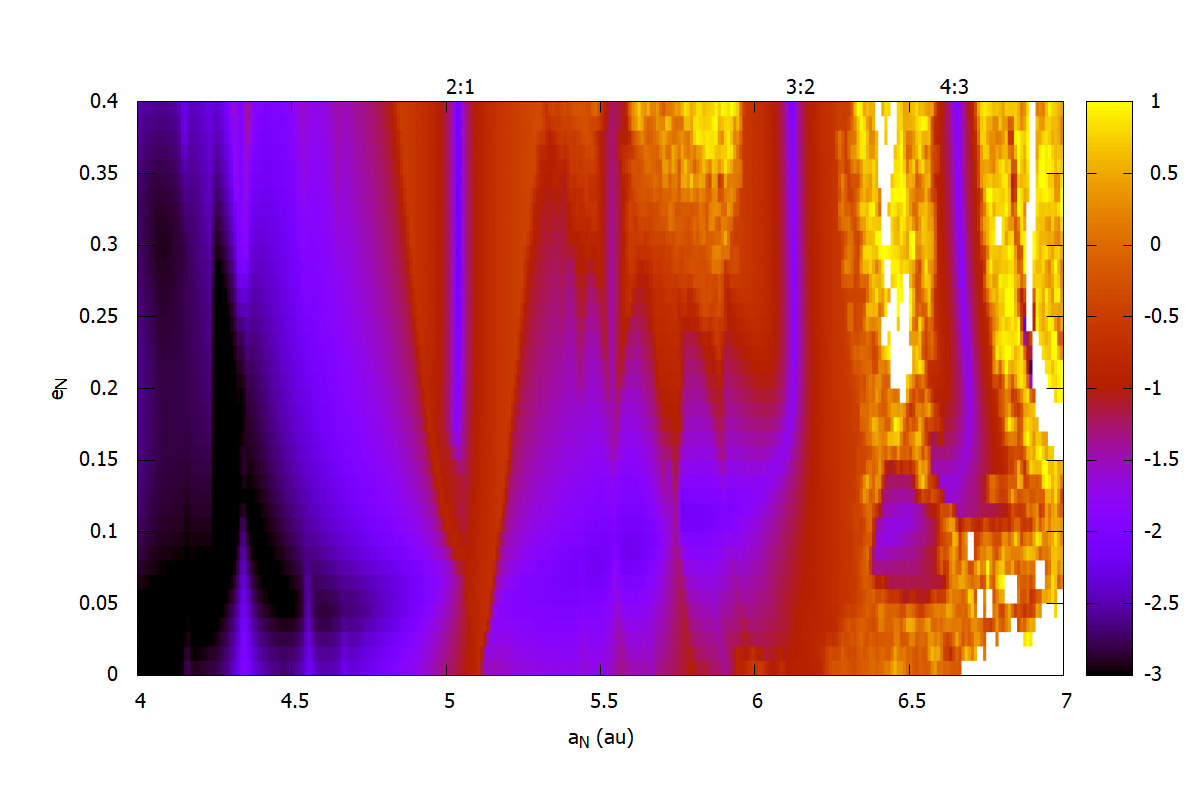}
	\caption{Dynamical map of a small region of Fig. \ref{atlasae} for the same initial conditions for the case  $\Delta\varpi=0\degree$  with  initial $\lambda_N=\lambda_J=0\degree$  obtained taking a grid of initial $(a_N,e_N)$ showing in color code scale maximum observed variations  $\log_{10}(\Delta \bar{a}_N)$ after 200 orbital revolutions. Black and blue colors correspond to small $\Delta \bar{a}_N$ corresponding to secular evolutions or evolution inside the central region of  MMRs, red corresponds to large amplitude librations  and yellow and white to very large $\Delta \bar{a}_N$ of unstable evolution.}
	\label{dmapae}
\end{figure}

In Fig. \ref{dmapae} we present a dynamical map calculated integrating numerically the exact equations of motion (no semi-analytical model involved) for a small region of Fig. \ref{atlasae} just to show that 
the atlas are a good indication of the actual dynamics. The code used for the numerical integrations was an adaptation of EVORB \citep{2002Icar..159..358F}. In color code logarithmic scale we show the maximum variation $\Delta \bar{a}_N$ obtained after 200 orbital revolutions of Neptune and were $\bar{a}_N$ refers to the osculating values filtered from short period oscillations.
The final appearance of the dynamical maps depend on the initial values for the complete set of orbital elements, in particular the mean anomalies which define the initial value of the resonant critical angle, but the general aspect is more or less replicated.
It is possible to distinguish in blue the deep central regions of resonances 2:1 (5.04 au), 3:2 (6.1 au) and 4:3 (6.6 au), the unstable regions at the right due to the proximity of Jupiter and the domain for the secular evolutions at the left part of the map. Resonances 3:2  and 4:3  show stable low $\Delta a_N$ evolutions inside a chaotic region at least in the short timescale adopted for the calculation of the dynamical map.  With its center at 5.04 au the wide 2:1 resonance dominates surrounded by a region of secular dynamics. The dynamical map has a richest structure because it is affected by short period and secular effects which the atlases do not, apart from simplifications inherent to the model.
Planetary resonances, contrary to the restricted case, generate libration amplitudes $\Delta a_i$ in both planetary semi-major axes. It is easy to show following \cite{Gallardo2021} that it is verified:
\begin{equation}
	\frac{\Delta a_1}{\Delta a_2} \simeq \frac{m_2}{m_1}(\frac{a_1}{a_2})^{2}
\end{equation}
then there is also a libration $\Delta a_J$ that we do not represent in the atlases (just $\Delta a_N$ are represented) and that have some effect 
in the dynamical map of Fig. \ref{dmapae} and affect its final aspect making resonances something wider than predicted in Fig. \ref{atlasae}. In any case, although atlases are ideal structures derived from a model, they constitute a useful reference, they do not depend on quick variables and are calculated quite easily. In the next subsection we will explore whether or not the overlap of resonances generates a chaotic regime.

\subsection{Regular and chaotic regimes}
\begin{figure}[hbt!]
	\centering   
	 \includegraphics[width=0.8\linewidth]{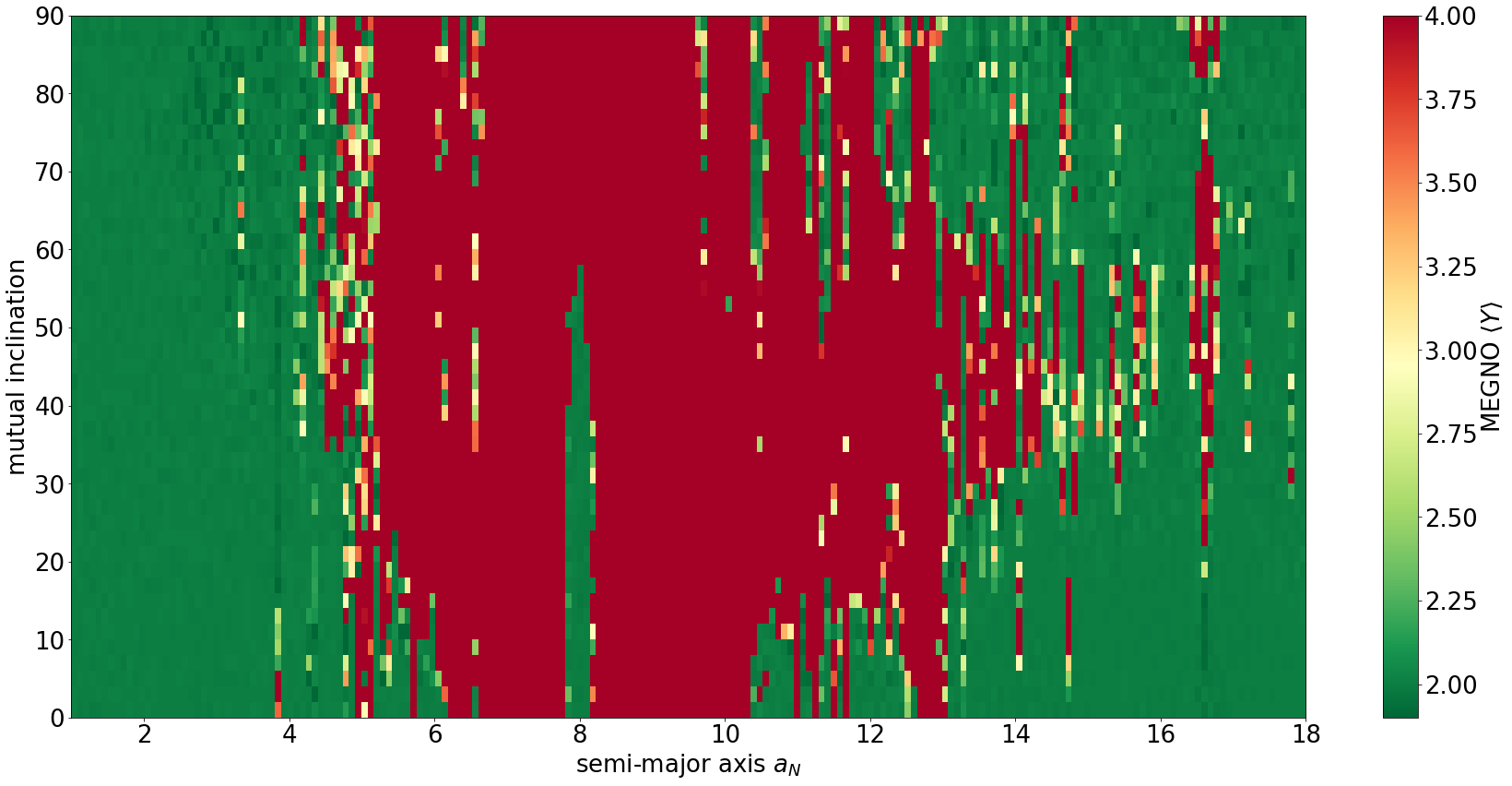}
	 \includegraphics[width=0.8\linewidth]{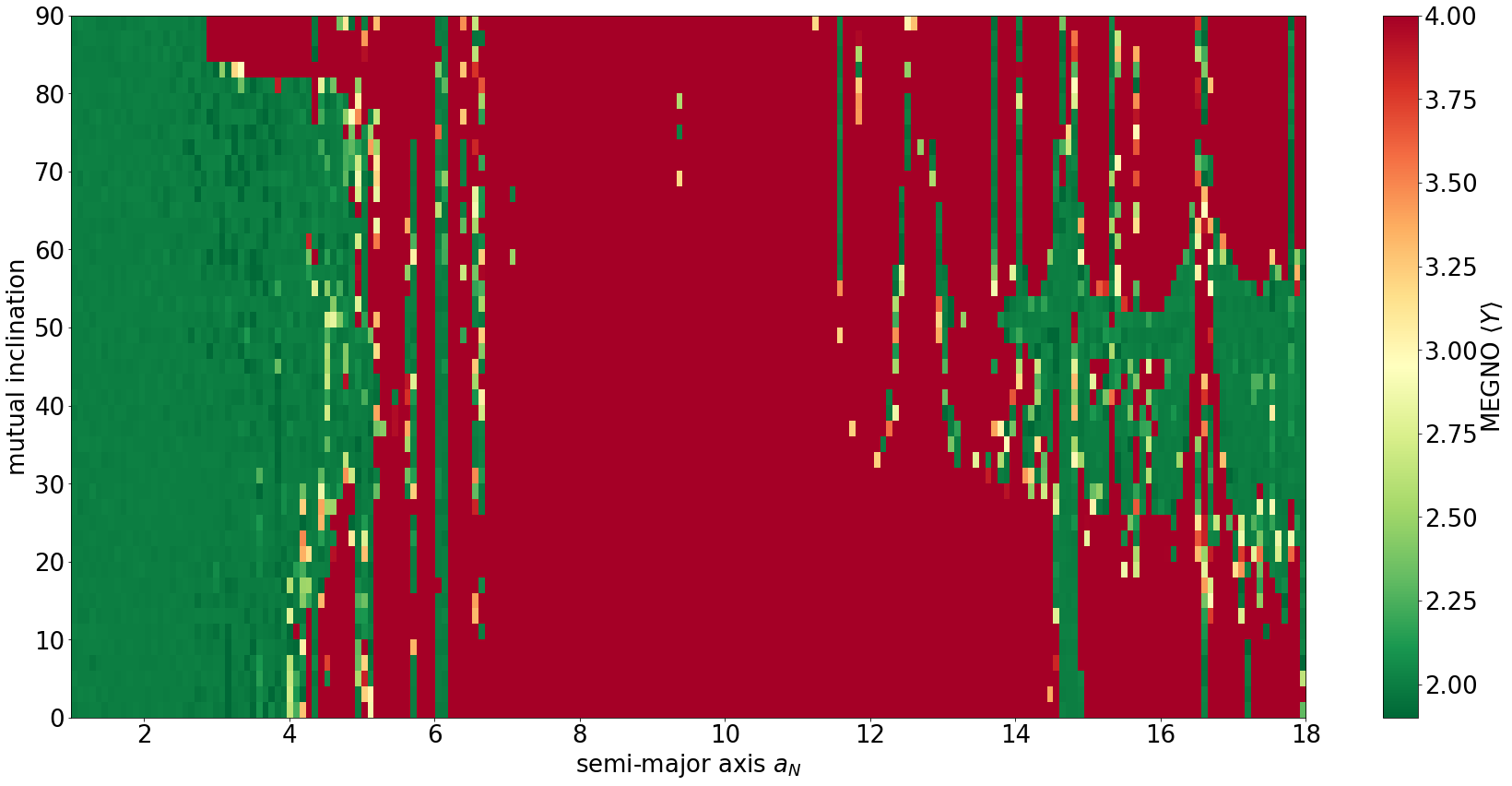}
	\caption{MEGNO indicator of chaos in the plane $(a_N,i)$ for a system of two planets with masses $m_J=0.001$  and $m_N=m_J/10$, with $a_J=8$ au and eccentricities $e_i=0.2$. Upper plot corresponds to $\Delta\varpi=0\degree$ 
		and lower plot to  $\Delta\varpi=180\degree$. In both plots the initial conditions are $\Omega_J=M_J=0\degree$, $\omega_J=90\degree$ and $\Omega_N=180\degree$. For the upper plot we took $\omega_N=270\degree$ and $M_N=60\degree$ and for the bottom one 
		$\omega_N=90\degree$ and $M_N=120\degree$.
		Green indicates regular dynamics and red chaotic. When the indicator is greater than 4 the value 4 was adopted. 
		The calculation time for computing MEGNO was 5000 orbital revolutions of the testing Neptune.}
	\label{megno}
\end{figure}

According to their orbital elements and masses it is expected that for some configurations planetary systems exhibit chaotic behavior.  In particular the mutual inclination seems to be a relevant parameter \citep{2011A&A...526A..98F,Libert2012,bhaskar2024}. As a basic exploration we show in Fig. \ref{megno} the computed MEGNO indicator of chaos \citep{2000A&AS..147..205C} for our working system in the space $(a_N,i)$ assuming $e_i=0.2$ for two situations: $\Delta\varpi=0\degree$ upper panel  and  $\Delta\varpi=180\degree$ in lower panel.  We use the code provided by REBOUND with the integrator ias15 \citep{2015MNRAS.446.1424R}
and with a total time span of 5000 orbital revolutions of the test Neptune. The regular regions are the green ones and the chaotic the red ones.

The time span could be not enough to detect chaos for some regions so some green regions could be chaotic in longer time spans, but the red ones are certainly chaotic, at least according to this simulation.
As far as MMRs are concerned the MEGNO map, like the dynamical map, is sensible to the initial $M_i$ and stable resonances can be showed as chaotic due to an inadequate choice of initial values. Resonance 1:1 is clearly showing regular dynamics for the case $\Delta \varpi=0\degree$.
We must remember that chaos is not equivalent to instability, it is possible to evolve chaotically but confined to a small region in the phase space.

Note the differences generated by the different initial $\Delta \varpi$, there is much more regular dynamics for $\Delta \varpi=0\degree$  and for the interior Neptune case.  The bottom panel can be compared directly with Fig. \ref{atlasai} and looking at this panel it is possible to identify all the chaotic region below $30\degree$ with the resonance overlap in Fig. \ref{atlasai}. Also, all the vertical chaotic structure from $0\degree$ to $90\degree$ between approximately 5 au and 13 au match very well with resonances overlap.
 Nevertheless the chaotic region for $i>80\degree$ at $a_N<5$ au and for  $i>50\degree$ at $a_N>14$ au certainly is not related to MMRs.
In this case other mechanism must be involved, which will be explored in the next section.

\section{Secular dynamics}

In this section we will explore the different secular regimes of our working system for a wide diversity of orbital configurations avoiding the MMRs mapped in the previous section.  First we will introduce the dynamical framework.

\subsection{Variables and equations}

The fundamentals of the reference system of variables for this problem can be found in  \citet{2006Icar..181..555M}, we include here  some fundamental definitions just for completeness.
In terms of the canonical Delaunay variables
\begin{equation}
(M, \omega, \Omega, L=\beta\sqrt{\mu a}, G=L\sqrt{1-e^2}, H=G\cos{i})
\end{equation}
the Hamiltonian of a star plus two planets is
\begin{equation}
	\mathcal{H}=-\frac{\mu_1^2\beta_1^3}{2L_1^2} -\frac{\mu_2^2\beta_2^3}{2L_2^2} -R(L_i,G_i,H_i,M_i,\omega_i,\Omega_i)
\end{equation}
being 
$R$ the disturbing function with $\mu_i=k^2(m_0+m_i)$, $\beta_i=m_0m_i/(m_0+m_i)$ and $k$ the Gauss gravitational constant.
In order to be canonical we must use the Jacobi or Poincare reference system \citep{2003ApJ...592.1201L,2006Icar..181..555M} and we will follow the last one.
Assuming there are no MMRs between the planets and excluding close encounters situations we can perform a method of mean eliminating the quick varying mean anomalies $M_i$ and we obtain the new, secular, Hamiltonian

\begin{equation}
	\mathcal{H}_s=-\frac{\mu_1^2\beta_1^3}{2L_1^2} -\frac{\mu_2^2\beta_2^3}{2L_2^2} -R_s(L_i,G_i,H_i,\omega_i,\Omega_i)
\end{equation}
where $R_s$ is the secular disturbing function which is proportional to $m_1m_2$ and obtained by a perturbation method or just by a numerical double averaging in $M_1$ and $M_2$. 
As $\mathcal{H}_s$ does not depend on $M_i$ we have that the new $L_i$ are constant (that means $a_i$ constant) and we can ignore them from the Hamiltonian, so the relevant part of $\mathcal{H}_s$ is just $R_s$.
Now we perform the canonical transformation defined by $(\omega_i,G_i,\Omega_i,H_i)$ $\longrightarrow$ $(\omega_i,G_i,\Omega_1,(H_1+H_2),(\Omega_2-\Omega_1),H_2)$. It results that, by  azimutal  invariance, $R_s$ is independent of  $\Omega_1$, then the momentum $(H_1+H_2)$ is constant and equal to the z component of the total angular momentum of the system.
Then the canonical equations for the secular motion of the two planets become:

\begin{equation}
	\frac{dG_1}{dt}=\frac{\partial R_s}{\partial \omega_1},  \hspace{1cm} \frac{d\omega_1}{dt}=-\frac{\partial R_s}{\partial G_1}   
\end{equation}
\begin{equation}
	\frac{dG_2}{dt}=\frac{\partial R_s}{\partial \omega_2},  \hspace{1cm} \frac{d\omega_2}{dt}=-\frac{\partial R_s}{\partial G_2}   
\end{equation}

\begin{equation}
	\frac{dJ}{dt}=0,  \hspace{1cm} \frac{d\Omega_1}{dt}=-\frac{\partial R_s}{\partial J}   
\end{equation}

\begin{equation}
	\frac{dH_2}{dt}=\frac{\partial R_s}{\partial \theta},  \hspace{1cm} \frac{d\theta}{dt}=-\frac{\partial R_s}{\partial H_2}   
\end{equation}
Where we define $J=(H_1+H_2)$ and $\theta=(\Omega_2-\Omega_1)=\Delta \Omega$. As $\Omega_1$ is not present in the Hamiltonian, the third equation can be ignored.
Note that the total angular momentum of the system must be constant and if we take its direction as z axis of our reference system (which is equivalent to take the invariable Laplace plane as reference plane) then the angular momentum vector for each planet must be opposed with respect to the z axis and this results in  $\Delta \Omega = 180\degree$ always, so the ascending nodes will be always opposed and evolving with the same time evolution. As the nodes are always opposed, just one of them is relevant for the dynamics. 
With $d\theta/dt=0$  the last equation becomes irrelevant also, but
the fact that $R_s$ is independent of $H_2$ does not imply that  $H_2$ is constant. 
Then, the resulting system 
can be reduced to  two degrees of freedom \citep{2006Icar..181..555M}, but if we analyze the time evolution of the orbital elements referred to an inertial frame we will find not two but three fundamental frequencies, assuming a regular motion.
These three fundamental frequencies are
$g_1$ and $g_2$, both associated to the longitude of the pericenters and eccentricities, and other one, $f$, associated with the common line of nodes and inclinations. 
These frequencies and some of their combinations will appear in the temporal evolution of the orbital elements for each planet and our main study is based on their analysis. An analytical deduction of these frequencies can be found in \cite{2007thesislibert} whenever one has an analytical expression for the Hamiltonian.

\subsection{$R_s$ topology and equilibrium configurations}

\begin{figure}[hbt!]
	\centering
	\includegraphics[width=1\linewidth]{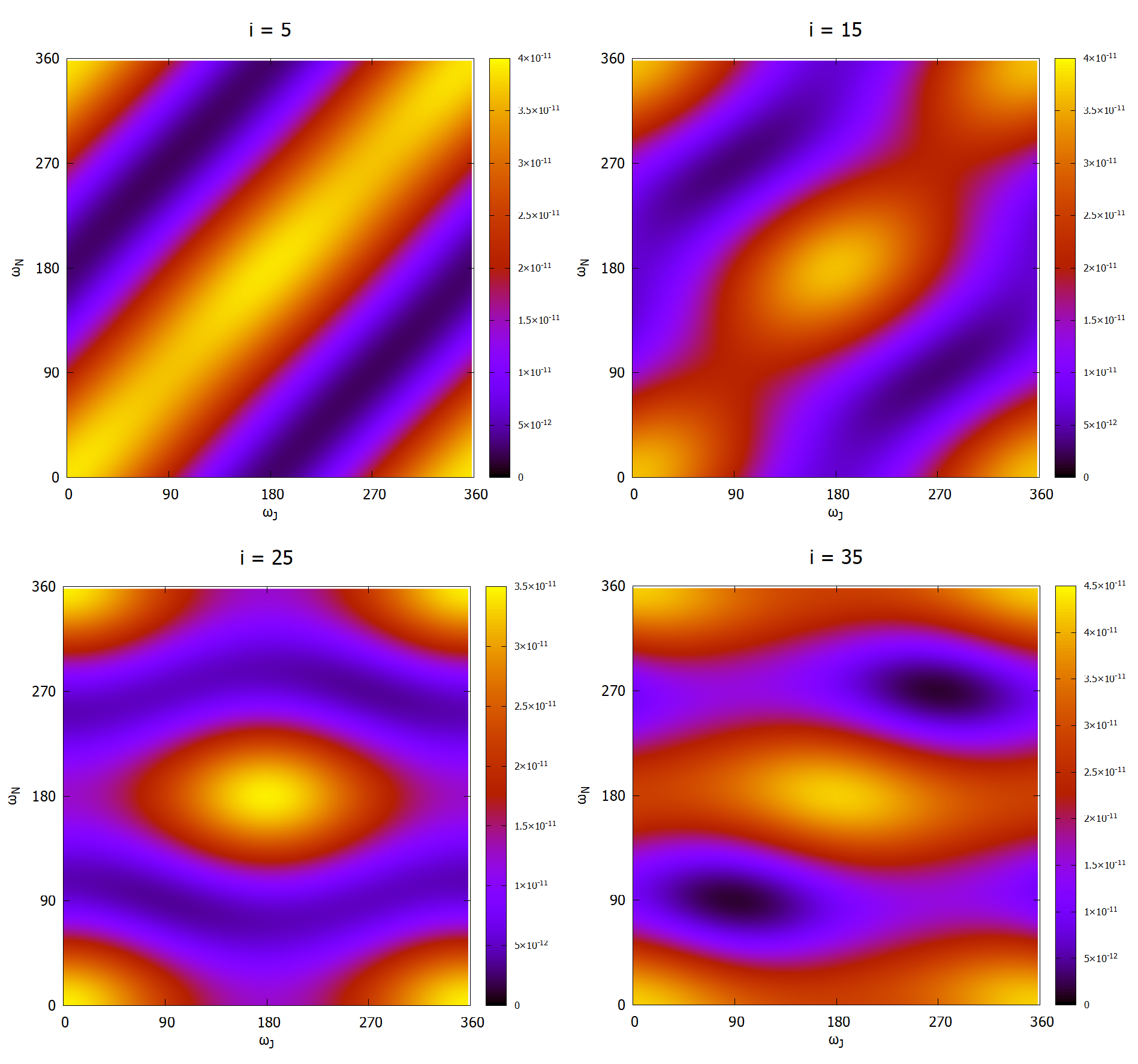}
	\caption{In color code is showed	$R_s-R_s(min)$ calculated numerically using Eq. (\ref{rsecu})  with  $a_N=4$ au,  $a_J=8$ au, $e_i=0.05$, $\Delta\Omega=180\degree$ and relative inclination of $5\degree$, $15\degree$, $25\degree$ and $35\degree$. Dark regions correspond to minima  and light regions to maxima. For $i>35\degree$ behavior is analogue to the last panel.} 
	\label{Rww}
\end{figure}

\begin{figure}[hbt!]
	\centering
	\includegraphics[width=1\linewidth]{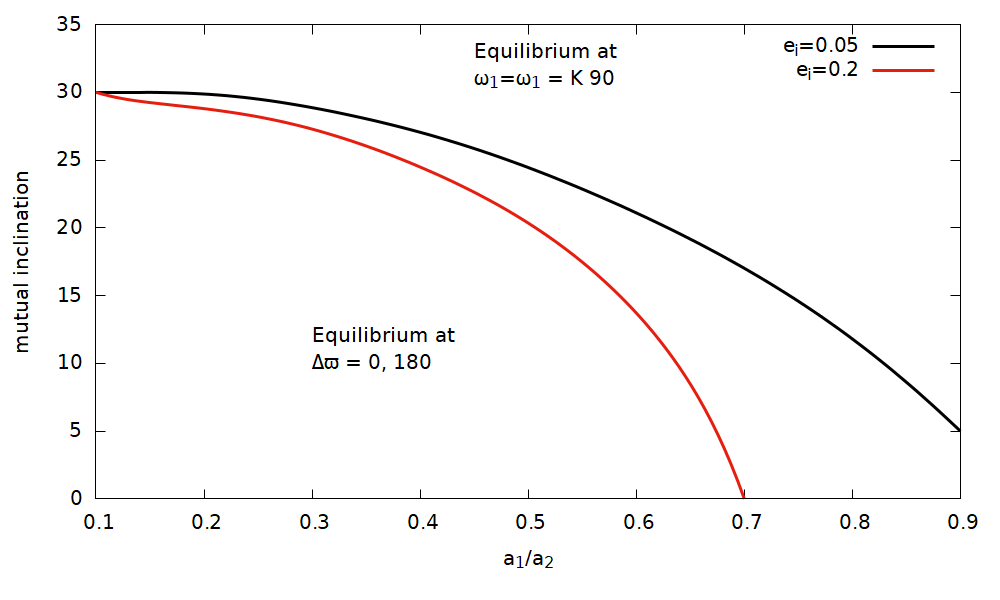}
	\caption{	Relative minimal mutual inclination for the onset of the extreme of $R_s$ at $\omega_1=\omega_2=\textsc{K} 90\degree$ as function of $a_1/a_2$ for two cases of eccentricity regimes: $e_i=0.05$ and  $e_i=0.2$.   Immediately below the curves there is a transition region and for lower inclinations the equilibrium occurs for $\Delta\varpi=0\degree,180\degree$.}
	\label{adependence}
\end{figure}

Taking the invariable Laplace plane as reference plane the spatial orientation of the mutual line of nodes does not have any physical meaning, then it has no dynamical relevance for the time evolution of the system. On the contrary, the mutual inclination and spatial orientations of the lines of apses given by $\omega_i$ are determinant for the dynamical evolution. To evaluate the proximity between both lines of apses for low inclination orbits  it is usual to consider $\Delta \varpi$ which in our case is equal to $\Delta \omega + 180\degree$, considering that $\Delta\Omega \equiv 180\degree$. There is a particular symmetry in the spatial configuration of the orbits when $\omega_i=0\degree, 180\degree$ or  $\omega_i=90\degree, 270\degree$ in the sense that there exists a plane of symmetry for the pair of orbits containing both line of apses. We can guess these values probably are associated to equilibrium configurations, and in fact previous works have showed that \citep{2006Icar..181..555M,2009MNRAS.395.1777M,Hansen2020}. We can explore the location of the equilibrium configurations in the subspace  $(\omega_1, \omega_2)$ 
calculating $R_s(\omega_1, \omega_2)$ by a numerical double averaging defined by
\begin{equation}
	R_s(\omega_1, \omega_2)=\frac{1}{4\pi^2}\int_0^{2\pi}\int_0^{2\pi} R(\omega_1, \omega_2) dM_1dM_2
	\label{rsecu}
\end{equation}
taking all other variables fixed (in particular  $\Delta\Omega \equiv 180\degree$).
Minima and maxima of $R_s(\omega_1, \omega_2)$ indicate the values of $\omega_i$ for equilibrium because they correspond to $\partial R_s/\partial \omega_i=0$ and then according to the canonical equations we have  $dG_i/dt=0$ which is a condition for equilibrium, although our analysis does not allow us to define whether they are stable or not. We should call them pseudo-equilibrium because we are restricted to the subspace of  $\omega_i$.
 Note that in the reference frame we are working the mutual inclination is $i=i_1+i_2$. 
Fig. \ref{Rww} shows the result of the calculation of Eq. (\ref{rsecu}) for planets with $e_i=0.05$  with four values of mutual inclination.
The minima and maxima of $R_s$ correspond to $\Delta\omega=180\degree$ and $0\degree$ respectively for the low-$i$ case (first panel) regardless the individual value of $\omega_i$. Considering that 
$\Delta \Omega \equiv 180\degree$, these extreme correspond to $\Delta\varpi=0\degree,180\degree$ which are known equilibrium configurations for the coplanar case. But, Fig. \ref{Rww} last panel shows the result for high-$i$ case and now the extremes are restricted to 
$\omega_1=\omega_2=\textsc{K} 90\degree$ where $\textsc{K}$ is an integer. For greater mutual inclination  the extreme remain there. There is a gradual transition between both regimes that in this example occurs for $15\degree<i<30\degree$ approximately and showed in the second and third panel of Fig. \ref{Rww}. We found a similar behavior in the pattern of the figures for other $e_i$ and $a_i$, but with the switch between the equilibrium points occurring at different inclinations, in particular they are strongly sensible to semi-major axes ratio.  These results are independent of the planetary masses because it is a purely geometric property considering that $m_i$ are just multiplicative factors in $R_s$, so we prepared the Fig. 
\ref{adependence} where the dependence with $a_1/a_2$ for two eccentricity regimes is showed, regardless the planetary masses. These curves define the minimum mutual inclination for the installation of the equilibrium points of $R_s$ at $\omega_1=\omega_2=\textsc{K} 90\degree$. Immediately below the curves there is the transition region like the  showed at the second and third panel of Fig. 
\ref{Rww} and for lower inclinations the equilibrium occur for $\Delta\varpi=0\degree,180\degree$. We must be careful with the results presented in Fig. 
\ref{adependence} because they were obtained using a secular model and we know that, according to the chaos map in Fig. \ref{megno}, for the case of two giant planets the region $a_1/a_2\gtrsim 0.6$ is chaotic, therefore the results are valid up to that value.

With our very simplistic analysis we cannot assure the system will evolve following lines of equal $R_s$ in Fig. \ref{Rww} because we assumed  fixed  $(e_i,i)$, but if the equilibrium points exist they should be at these extreme in the plane $(\omega_1,\omega_2)$. These equilibrium configurations are independent of the masses because they are purely geometric consequence but the orbital time evolution depend on the planetary masses.
In general, we observed that $\omega_N$ tend to oscillate (vZLK mechanism) but $\omega_J$ circulates for small $m_N/m_J$ ratios but for greater $m_N$ both $\omega_i$ can oscillate.
We will take some of these combinations of $\omega_i$ for the initial conditions of our numerical calculation of the fundamental frequencies. An analytical justification of the equilibrium configurations can be found in \cite{2009MNRAS.395.1777M}.

\begin{figure}[hbt!]
	\centering
	\includegraphics[width=1\linewidth]{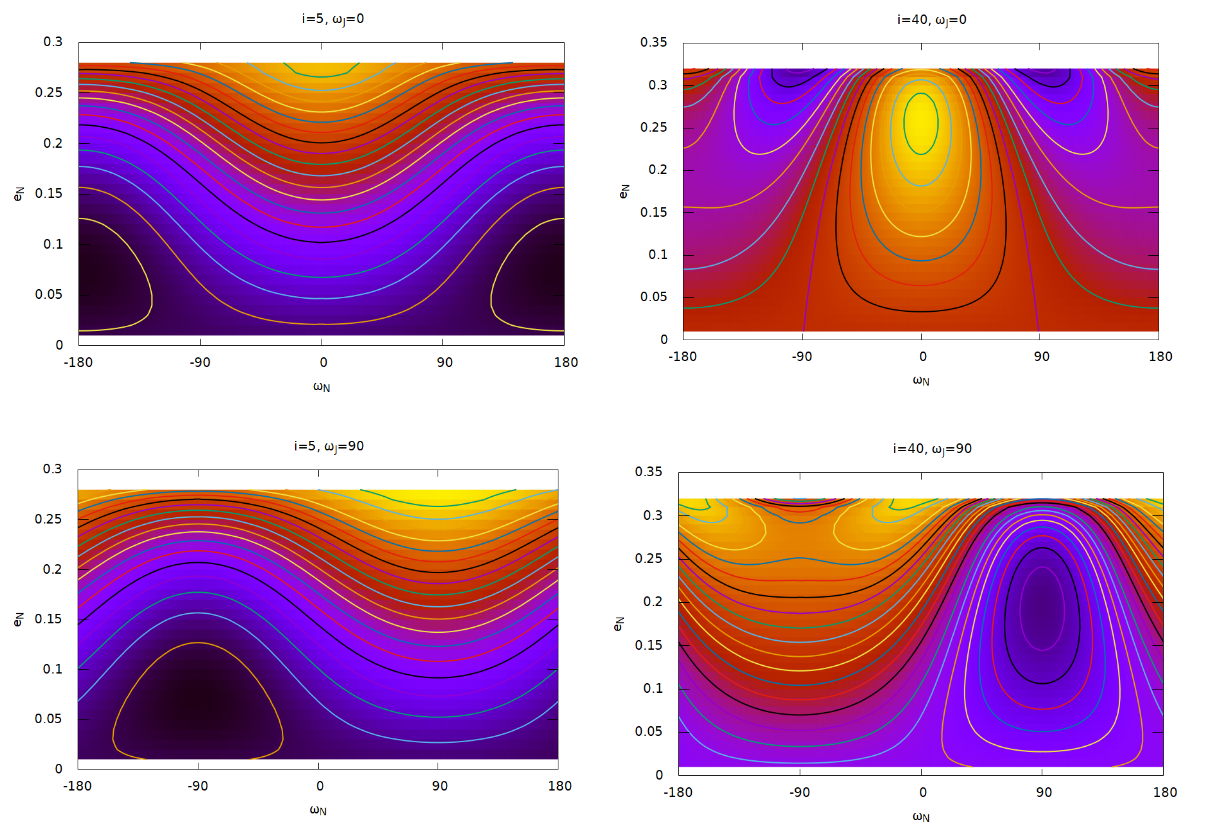}
	\caption{Level curves of $R_s$ showed in color code (black for lower values and yellow for high values) and  location of the pseudo equilibrium points in the plane  $(\omega_N, e_N)$ for fixed values of $(\omega_J, i)$ varying $e_J$ conserving the total angular momentum of the system, defined by the initial $e_J=e_N=0.1$ and the assumed mutual inclination. Level curves of $R_s$ for low inclination orbits ($i=5\degree$) at left and high inclination orbits ($i=40\degree$) at right. Top panels for $\omega_J=0\degree$ and bottom panels for $\omega_J=90\degree$. 
		Planets are at $a_J=8$ and $a_N=15$ au with $\Delta\Omega=180\degree$.}
	\label{Rwe}
\end{figure}

As we will explain in the next subsection, we noted that in the several examples we studied with different initial conditions the mutual inclination remains  almost constant for initial $i$ less than some limit value that we will identify with the beginning of the vZLK mechanism.
We take advantage of the fact that mutual inclination is almost constant and then we  assume it fixed to calculate level curves of $R_s(\omega_N, e_N)$  taking $e_J$  so that the total angular moment remains equal to some constant value, which is more realistic than assuming constant $e_i$. Fig. \ref{Rwe} shows four examples corresponding to two different relative inclination regimes and for two different initial $\omega_J$. As in Fig. \ref{Rww} dark colors represent minima for $R_s$  and light ones  maxima. The curves represent possible evolution paths as long as $i$ remains constant and $\omega_J$ is fixed at the initial value.

Left panels of Fig. \ref{Rwe} corresponding to $i=5\degree$ indicate that the  equilibrium points are at $\Delta\varpi=0\degree$ for low eccentricity and a point at $\Delta\varpi=180\degree$ is suggested for high eccentricities  and regardless of the individual $\omega_i$ because the topology is the same, just shifted in $\omega_N$. Note this is in agreement with Fig \ref{Rww} first panel and also with results already known for the coplanar case \citep{2003ApJ...592.1201L,2004Icar..168..237M}.
Right panels of Fig. \ref{Rwe} show the topology for $i=40\degree$ which are strongly dependent on $\omega_i$: one at $\omega_i=0\degree$  in the upper panel  and one at  $\omega_i=90\degree$  in the low panel  in agreement with Fig \ref{Rww} last panel.
Then, while in the low inclination case the equilibrium configuration occurs for $\Delta\varpi = 0\degree,180\degree$, in the high inclination case  the $\omega_i$ cannot be arbitrary but, as we have mentioned earlier,  multiples of $90\degree$. For low inclination orbits the line of apses can be in arbitrary position respect to the mutual line of nodes (just $\Delta \varpi$ matters) but for high inclination orbits the  equilibrium configurations correspond to line of apses coincident or perpendicular to the line of nodes.

\subsection{Fundamental frequencies and dynamical regimes}

Assuming a regular evolution for the system, the fundamental frequencies and some of their combinations will appear in the time evolution of the orbital elements. For low eccentricity and low inclination orbits they can be reproduced by the Lagrange-Laplace theory \citep{1961mcm..book.....B,1999ssd..book.....M} but for high $(e,i)$ orbits a different approach must be used. For example, a higher order analytical theory can be found in \cite{2008CeMDA.100..209L} and \cite{Libert2012}.
In our study, to obtain the fundamental frequencies we carry on numerical integrations of the exact equations of motion, typically for 10 Myr,  and examine the time evolution of the variables
 $(k_i,h_i)=e_i(\cos\varpi_i,\sin\varpi_i)$ and $(q_i,p_i)=i_i(\cos\Omega,\sin\Omega)$ being 
$i_i$ the inclination respect to the invariable plane. The output of the numerical integrations were analyzed  using the frequency analysis proposed by \cite{Ferraz-Mello1981} and \cite{Gallardo1997} and the code given by \cite{2017ascl.soft01003G}.
The method, which is efficient detecting low frequencies, calculates a series of frequencies presented in the time series, each one associated to a spectral correlation coefficient (SCC) which is always less than 1 and related to the probability that the time evolution is byproduct of a real periodicity with that frequency and not product of chance.
For low to middle $(e,i)$ we found the two frequencies $g_1$ and $g_2$  obtained from variables  $(k,h)$ and the frequency $f$ obtained from $(q,p)$, following the usual notation for the Solar System. At high $(e,i)$ regimes we found that these three frequencies can appear with its harmonics and combinations in all variables. For more extreme values of $(e,i)$  several frequencies appear in the spectra, indicating a more complex and chaotic behavior.  In these cases, identifying the fundamental frequencies is more difficult, so we simply record all those with  SCC$>0.05$.

\begin{figure}[hbt!]
	\centering
	\includegraphics[width=1\linewidth]{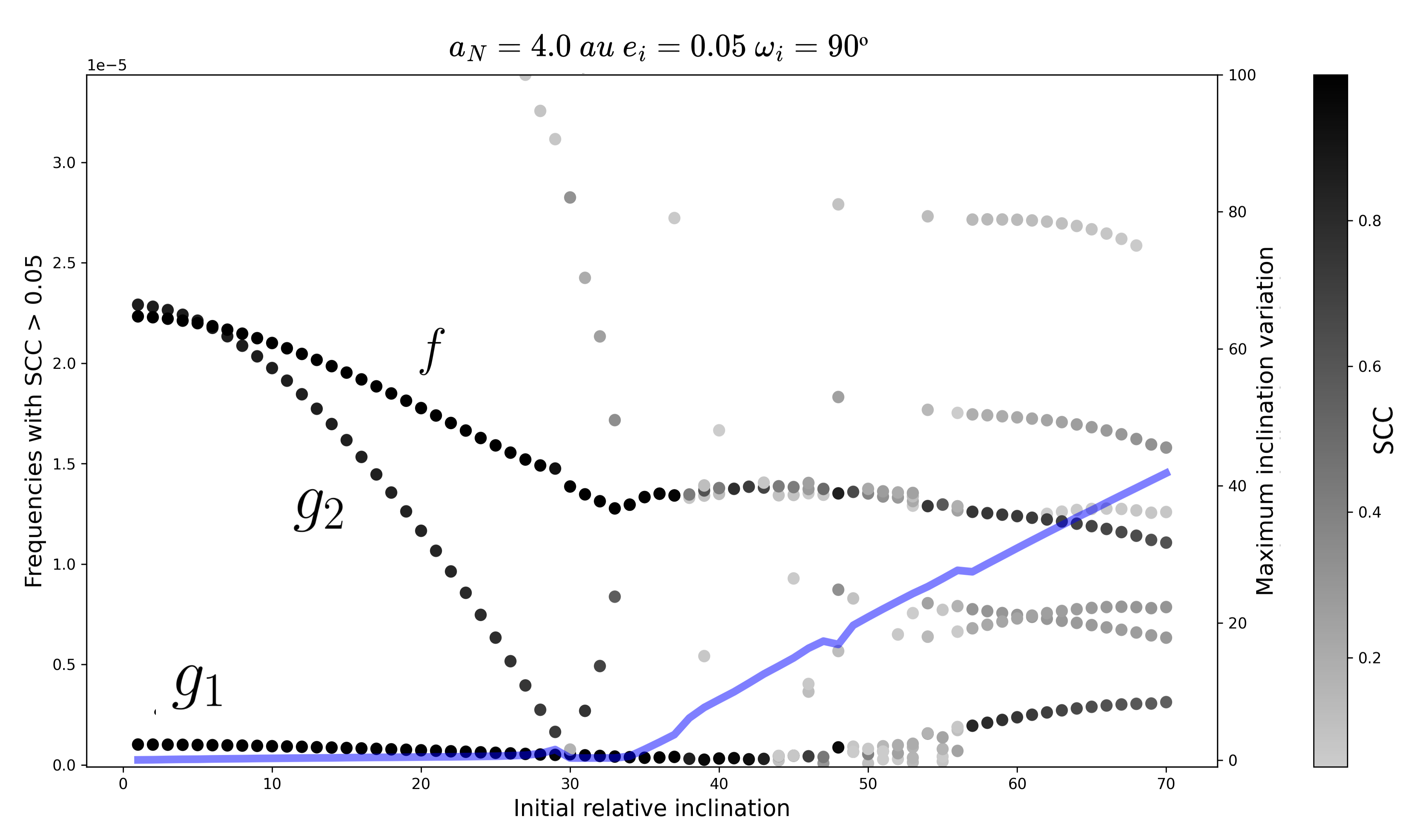}
	\caption{Frequencies in units of yr$^{-1}$ detected with SCC$>0.05$ in grayscale as a function of initial mutual inclination from numerical integrations over 10 Myr.  Blue curve indicates maximum variation in relative inclination. Initial values of $e$ and $\omega$ for both planets are indicated in the title.}
	\label{ffsample}
\end{figure}

In Fig. \ref{ffsample} we show one example of the dependence of the frequencies with the initial relative inclination between the planets for low initial eccentricities ($e_i=0.05$). Analyzing the output of the numerical integrations we obtain the spectra and we plot in gray scale all detected frequencies with spectral correlation coefficient SCC$>0.05$. Frequency $f$ corresponds to a retrogradation of the common line of nodes ($d\Omega/dt<0$) while the $g_i$ are related to the motion of the  pericenters $\varpi_i$
which are prograde for low inclinations.
For low $(e,i)$ regimes it is obtained the classical result for the asteroidal case were $f$ and the higher  $g_2$ (related to the lower mass planet) are very close while the low $g_1$ is related to $\varpi_J$,
the longitude of the pericenter of the more massive planet. 
For increasing mutual inclinations all frequencies diminish and, in the case of the figure, for $i\sim 30\degree$ it happens that  $g_2=g_1$ which is a condition for the establishment of  a secular resonance  which in case of low eccentricity regime it does not represent a destabilizing condition, at least in this example. The critical inclination, $i_c$, for the occurrence of $g_2=g_1$ in all experiments we performed for the range of $a_N$ we studied is located between $30\degree$ and $40\degree$. The value of $i_c$ seems to be not equal but linked to the change in the distribution of the equilibrium points of $R_s$ showed in Fig. \ref{Rww}.
For higher inclinations $g_2$
cross the zero and starts to grow but in the plane $(k,h)$  the sense of circulation is the contrary, that means retrograde. For $i\sim 36\degree$ occurs that $g_2=f$ which implies that $\Omega$ and $\varpi$ retrogradate with approximately the same frequency resulting in an oscillation of $\omega_N$, which characterizes the vZLK mechanism. At this point the mutual inclination starts to change more and more as greater  its initial value. In this example, the greater the relative inclination the more complex the motion. In the region $i>40\degree$ the  complexity of the dynamics is so great that only three pure frequencies cannot describe the system evolution and combinations between them may appear. Note that for $i>i_c$ the frequency $f$ is accompanied by the frequency $f + (f-g_2)$. Note also that after $g_2=f$ the frequency $g_2$ disappears or is merged with $f$  and the frequency $f$ shows a discontinuity defining a change in the dynamical regime: the installation of the vZLK regime. It is worth clarifying that although at some point at low inclinations $g_2=f$ occurs, this does not constitute a secular resonance since they actually have opposite signs and, in addition, for these inclinations the $(k,h)$ variables are decoupled from the $(q,p)$ variables.

\begin{figure}[hbt!]
	\centering
	\includegraphics[width=1\linewidth]{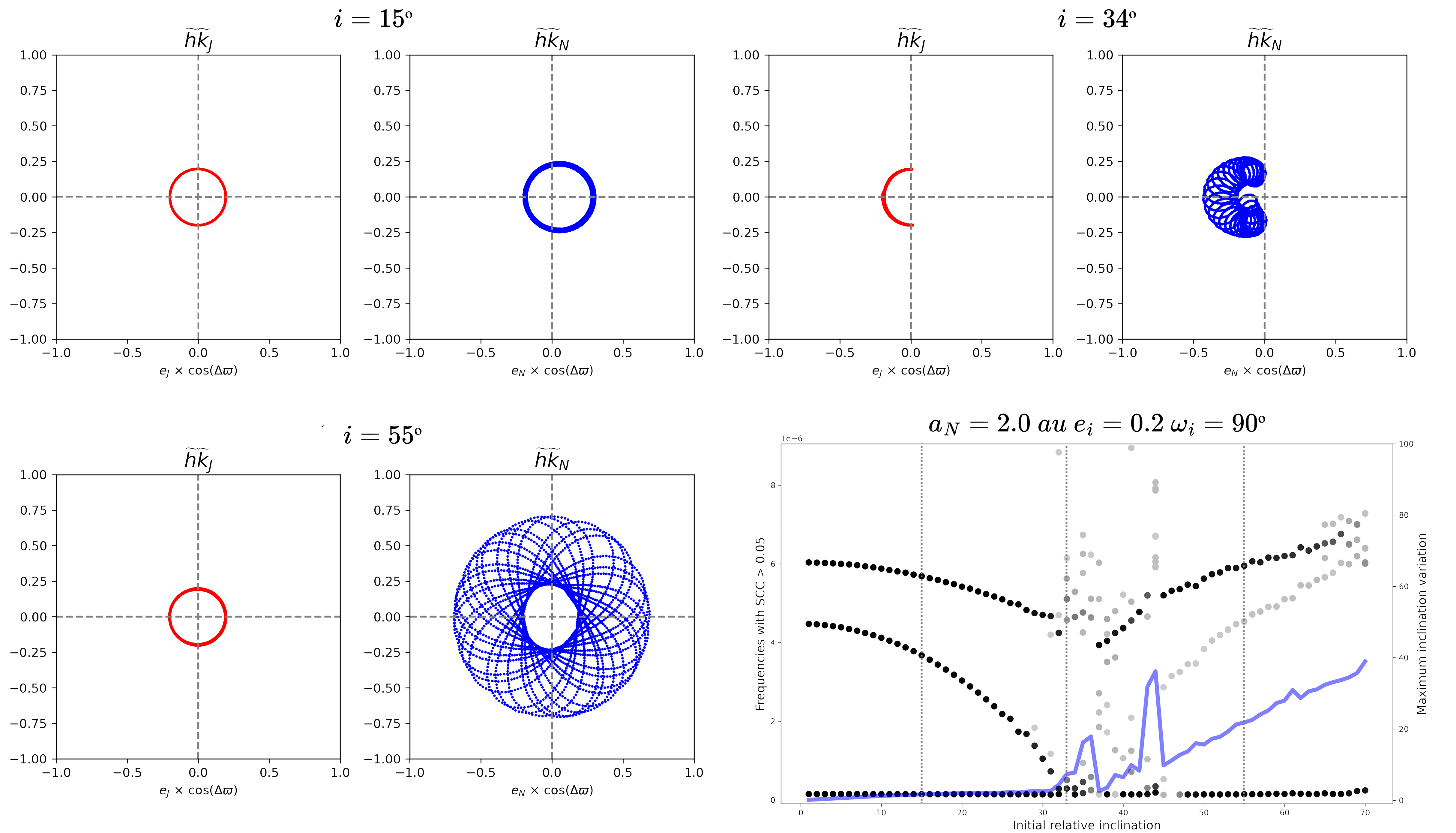}
	\caption{Three regimes for different intial mutual inclinations: classic secular ($i=15\degree$), secular resonance ($i=34\degree$) and vZLK mechanism ($i=55\degree$). The last panel shows the sequence of frequencies with SCC$>0.05$ and the vertical lines indicate the positions for the three previous panels.}
	\label{3cases}
\end{figure}

Examples of trajectories with different  evolution paths are showed in Fig. \ref{3cases}, in this case for more eccentric orbits ($e_i=0.2$). For low inclination regime (first panel), $i<i_c$, the behavior is analogue to the classic secular dynamics in the planes $(k,h)$ and $(q,p)$. 
In the figure we plot $(\tilde{k}=e_i\cos\Delta\varpi,\tilde{h}=e_i\sin\Delta\varpi)$ instead of $(k,h)$ because $\Delta\varpi$ is more relevant than the individuals $\varpi_i$. 
We will call this regime, where frequencies $g_i$ appear in $(k,h)$ and the frequency $f$ in $(q,p)$, as \textit{classic secular} and approximately corresponds  with what in \citet{2006Icar..181..555M} is called domain 4  and in \cite{2023CeMDA.135...22M} the 'planar-like' regime. 
For $i\sim i_c$ the secular resonance $g_2=g_1$ happens and banana like trajectories in $(\tilde{k_i},\tilde{h_i})$ show up (second panel). For $i>i_c$ a chaotic region appears and for higher inclinations a vZLK like mechanism starts with changes in mutual inclination which implies larger changes in eccentricities via angular momentum conservation which makes the variables $(k,h)$ and $(q,p)$ no longer decoupled.
 This regime corresponds approximately with domains 1 to 3 in \citet{2006Icar..181..555M}. For this region the plot $(\tilde{k},\tilde{h})$ in the third panel does not reveal the true librations of $\omega_N$ typical of vZLK but can be easily verified as other authors have shown.

It is important to stress that apsidal librations of $\Delta\varpi$ can be observed for low mutual inclinations and it is not because $g_2=g_1$. That is typical of several planar or quasi planar extrasolar systems and it is not due to a secular resonance but to a large forced mode that shift the circular trajectories away from the center in the plane  $(\tilde{k},\tilde{h})$. 
If the forced eccentricity in the first panel of Fig. \ref{3cases} were much larger, the blue circle would exclude the center of the plot and $\Delta\varpi$ would look like an oscillation around $0\degree$.
Instead, the secular resonance occurs by definition when the both fundamental frequencies are equal.

\begin{figure}[t]
	\centering
	\includegraphics[width=1\linewidth]{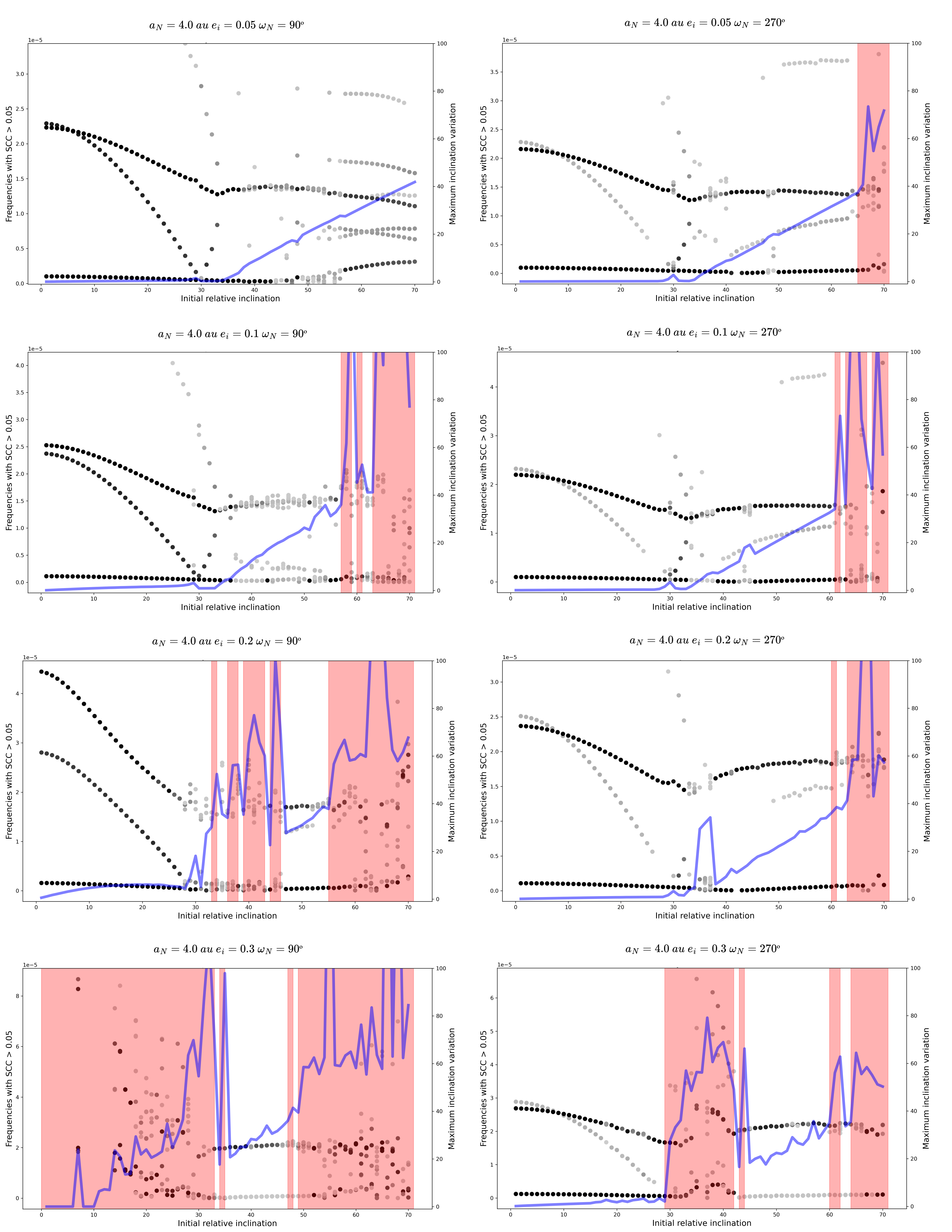}    
	\caption{Frequencies in units of yr$^{-1}$ detected with SCC$>0.05$ in grayscale as a function of initial mutual inclination from numerical integrations over 10 Myr and  in all cases with initial $\omega_J=90\degree$.
		The blue line indicates maximum changes in mutual inclination. Red regions correspond collision or to observed changes $\Delta a_i/a_i > 0.02$ indicating departure from secular regime.  
		Two initial configurations for $\omega_i$ are showed: left column corresponds to initial $\Delta\varpi=180\degree$ and right column to $\Delta\varpi=0\degree$. From top to bottom increasing eccentricities of both planets. All plots refers to an interior Neptune like planet with $a_N=4$ au.}
	\label{8cases1}
\end{figure}

In Fig. \ref{8cases1} we show a mosaic of figures like Fig. 
\ref{ffsample} corresponding to two initial different  configurations of $\omega_i$ (left and right) and for four regimes of orbital eccentricities from top to bottom. We indicate in red some initial conditions where we obtained collisions or variations $\Delta a_i/a_i > 0.02$ indicating that the secular evolution is compromised by strong perturbations.
For low $e$ there is no much difference between left and right columns, but for high $e$ the initial $\Delta\varpi=180\degree$ of left column generate more instabilities than the initial   $\Delta\varpi=0\degree$ of the right column. Frequency $f$ change with $i$ but never reaches zero, while $g_2$ always drops to zero or cross $g_1$ for some critical $i_c$. The frequency $g_1$ is always small and shows very little changes with $i$. Presence of 
 frequencies $f + (f-g_2)$ and the merging of $g_2$ with $f$ are also observed.

\begin{figure}[hbt!]
	\centering
	\includegraphics[width=1\linewidth]{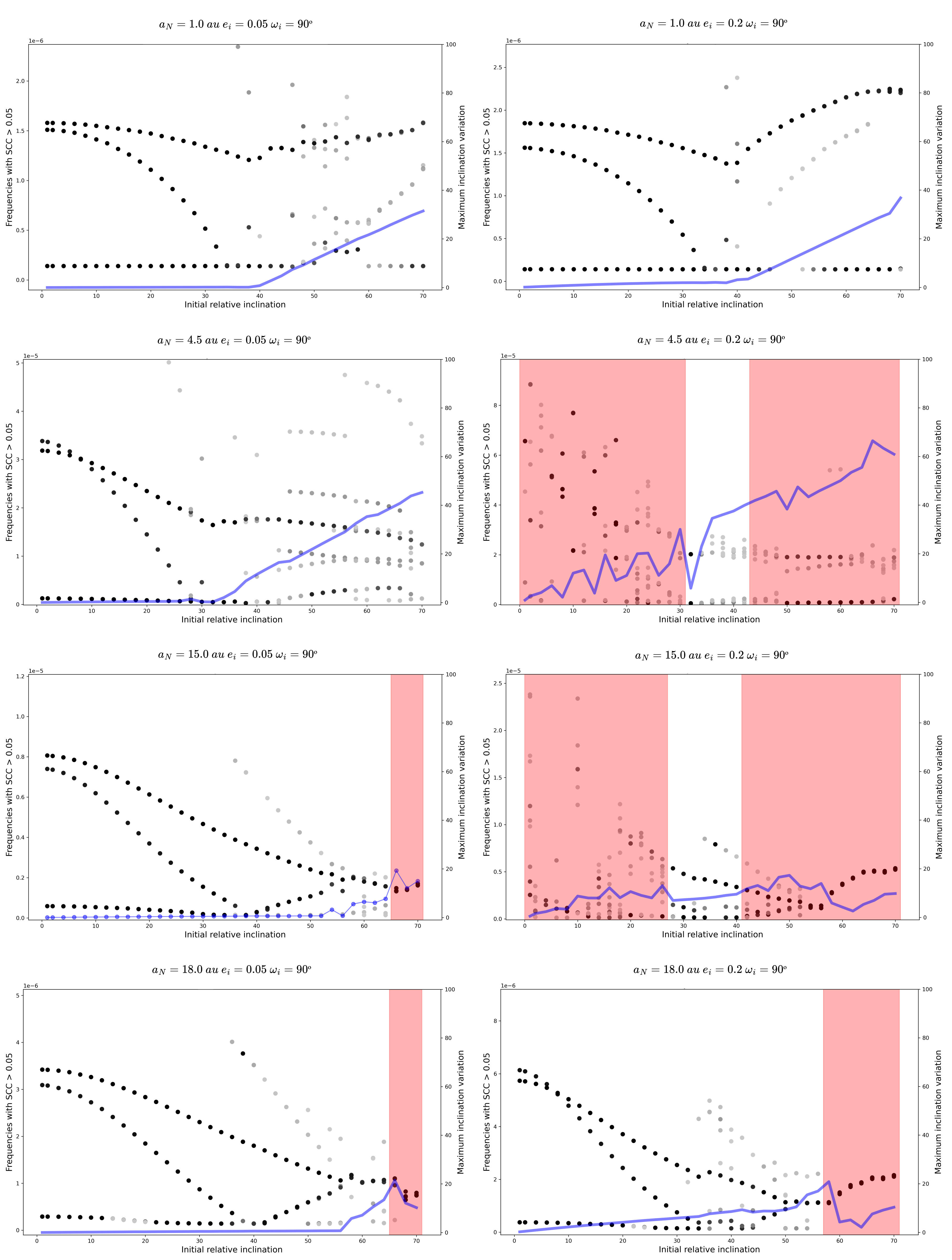}   
	\caption{Same as Fig. \ref{8cases1}  for initial $\omega_i=90\degree$. Left and right columns correspond to initial low ($e_i=0.05$) and high ($e_i=0.2$) eccentricity regimes respectively. From top to bottom increasing $a_N$. }
	\label{8cases2}
\end{figure}

The behavior of the fundamental frequencies for increasing $a_N$ is showed in Fig. \ref{8cases2} from top to bottom for the case of low  and high eccentricity orbits (left and right respectively).
In general, in all experiments with systems  verifying    $0.6 \lesssim a_N/a_J \lesssim 1.8$ for the eccentric case $e_i=0.2$  we obtained destruction of the secular behavior 
due to the proximity between planets.
The unstable configuration for $e_i=0.2$ and $a_N=4.5$  corresponds with the chaotic regions close to Jupiter showed in Fig. \ref{megno}. 
From the inspection of Fig. \ref{8cases2} is possible to deduce some effect of $a_N$ on $i_c$ and also on the value of $i$ corresponding to the beginning of the vZLK mechanism which for Neptune exterior to Jupiter occurs at a higher $i$.

\begin{figure}[hbt!]
	\centering
	\includegraphics[width=1\linewidth]{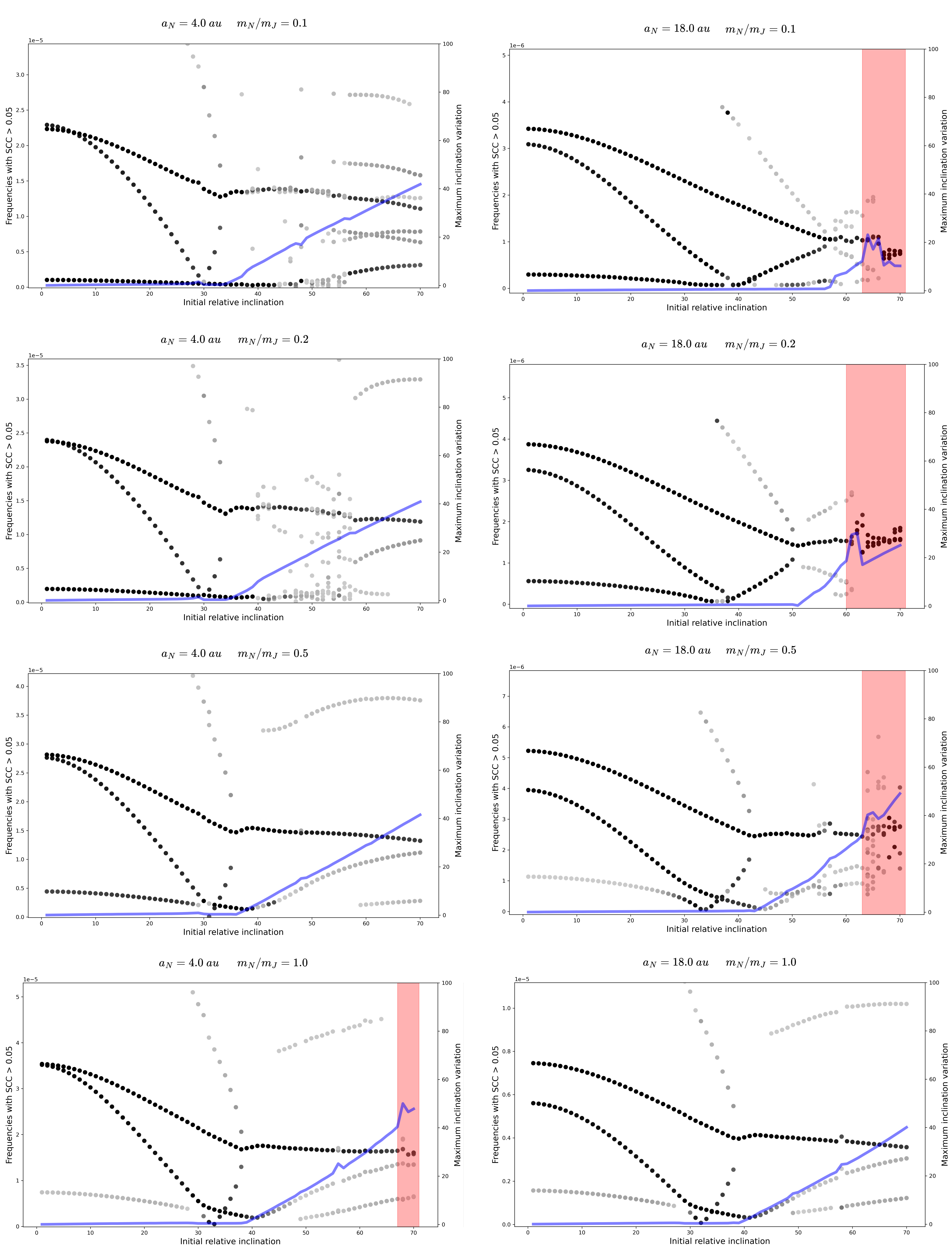}   
	\caption{Same as Fig. \ref{8cases1}  for initial $\omega_i=90\degree$ and low eccentricity regime $e_i=0.05$. Left and right columns correspond to $a_N=4$ and $a_N=18$ respectively. From top to bottom increasing $m_N/m_J$ ratio.}
	\label{8cases3}
\end{figure}

Finally we studied the effect of the ratio $m_N/m_J$  on the fundamental  frequencies. Fig. \ref{8cases3} shows the frequencies for four values of the mass ratio increasing $m_N$ from top to bottom for low eccentricity regimes and for two values of $a_N$ corresponding to an interior Neptune (left column) and an exterior one (right). For larger $m_N$ the frequency $g_1$ grows due to the increase of the perturbation of Neptune on Jupiter. In the right column corresponding to the exterior Neptune is very clear the shift of the vZLK mechanism to higher inclinations for diminishing values of $m_N$. We have also experiments with larger eccentricities but the instabilities are reached just after the secular resonance. Note that for larger $m_N$ things become more complex: sometimes $g_1$ is missing after the critical inclination, sometimes is $g_1$ not $g_2$ which drop to zero and bounces. This behavior is expected when the masses are comparable and the frequencies are not associated with a particular planet but with the system as a whole.

Apart from instabilities due to close encounters or captures in MMRs the general behavior is that, as long as the system maintains its stability,
all frequencies decrease for growing initial mutual inclinations up to some critical $i_c$ when the crossing of $g_2$ with $g_1$ stops the well behaved secular dynamics and a secular resonance arises. 
 For mutual $i>i_c$  the
system enters in a different more complex regime with changes in its mutual inclination that specifically start when $g_2=f$
 initiating the vZLK mechanism.
As we have explained earlier we noted that in the several examples we studied the mutual inclination remains almost constant for initial $i<i_c$ and we
 took advantage of this   for constructing Fig. \ref{Rwe}.

Studying trajectories in the plane $(\Tilde{k},\Tilde{h})= e_i(\cos{\Delta\varpi},\sin{\Delta\varpi})$ it is possible to identify a forced and a free eccentricity for each planet, that means a fixed vector (the forced one) around which another free vector evolves. The forced modes are well defined for low inclinations where the $g_i$ are different.
Fig. \ref{3cases} first panel shows the classical behavior: the less massive planet describes a circle with radius equal to the free eccentricity shifted from the center an amount equal to the forced eccentricity. For the massive planet the forced eccentricity is minimal (that makes its osculating eccentricity almost constant) but detectable.
When $g_2\sim g_1$ (second panel in Fig. \ref{3cases})
the concept of forced mode losses its significance. That means, close to the secular resonance  the behavior is very different and, as can be seen in Fig. \ref{megno},  the secular resonance is inside the regular region of the MEGNO map.  For $i> i_c$ a chaotic region arises and then, for greater inclinations  there is a regular region with the typical path given by the third panel in Fig. \ref{3cases} where the dynamics is characterized by the vZLK mechanism with oscillations of $\omega_N$.

\begin{figure}[hbt!]
	\centering
	\includegraphics[width=1\linewidth]{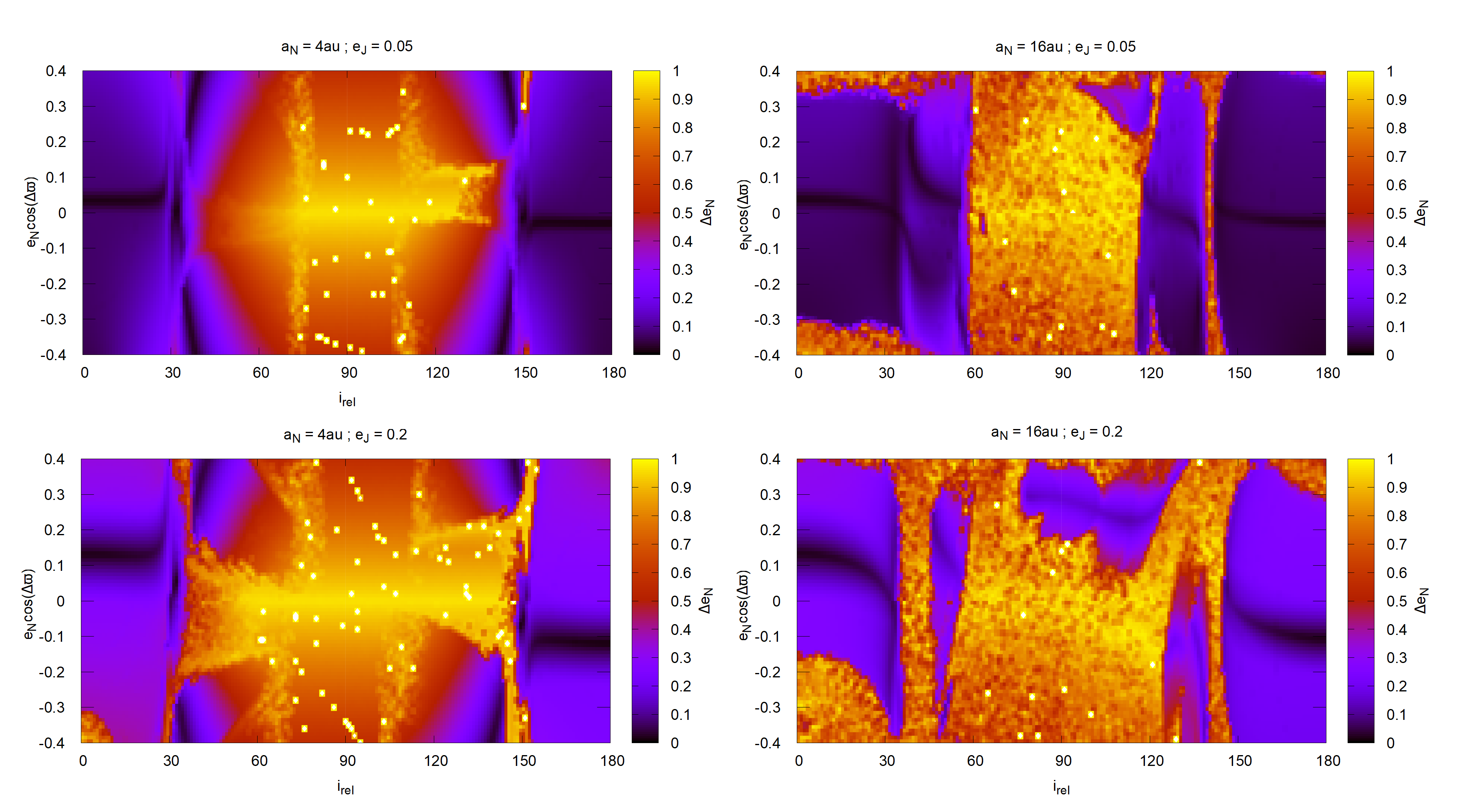}
	\caption{Dynamical maps showing changes $\Delta e_N$ for $a_N=4$ ua (left) and $a_N=16$ au (right) for $e_J=0.05$ (up) and 
		$e_J=0.2$ (below). In all panels the upper half corresponds to $\Delta\varpi=0\degree$ 	($\omega_J=90\degree$, $\omega_N=270\degree$) and the lower half to  $\Delta\varpi=180\degree$ ($\omega_i=90\degree$).
 Integration time $=10^5$ orbital revolutions. Note that the variations $\Delta e_J$ might be relevant, but are not shown here. }
	\label{deltae}
\end{figure}

\begin{figure}[hbt!]
	\centering
	\includegraphics[width=1\linewidth]{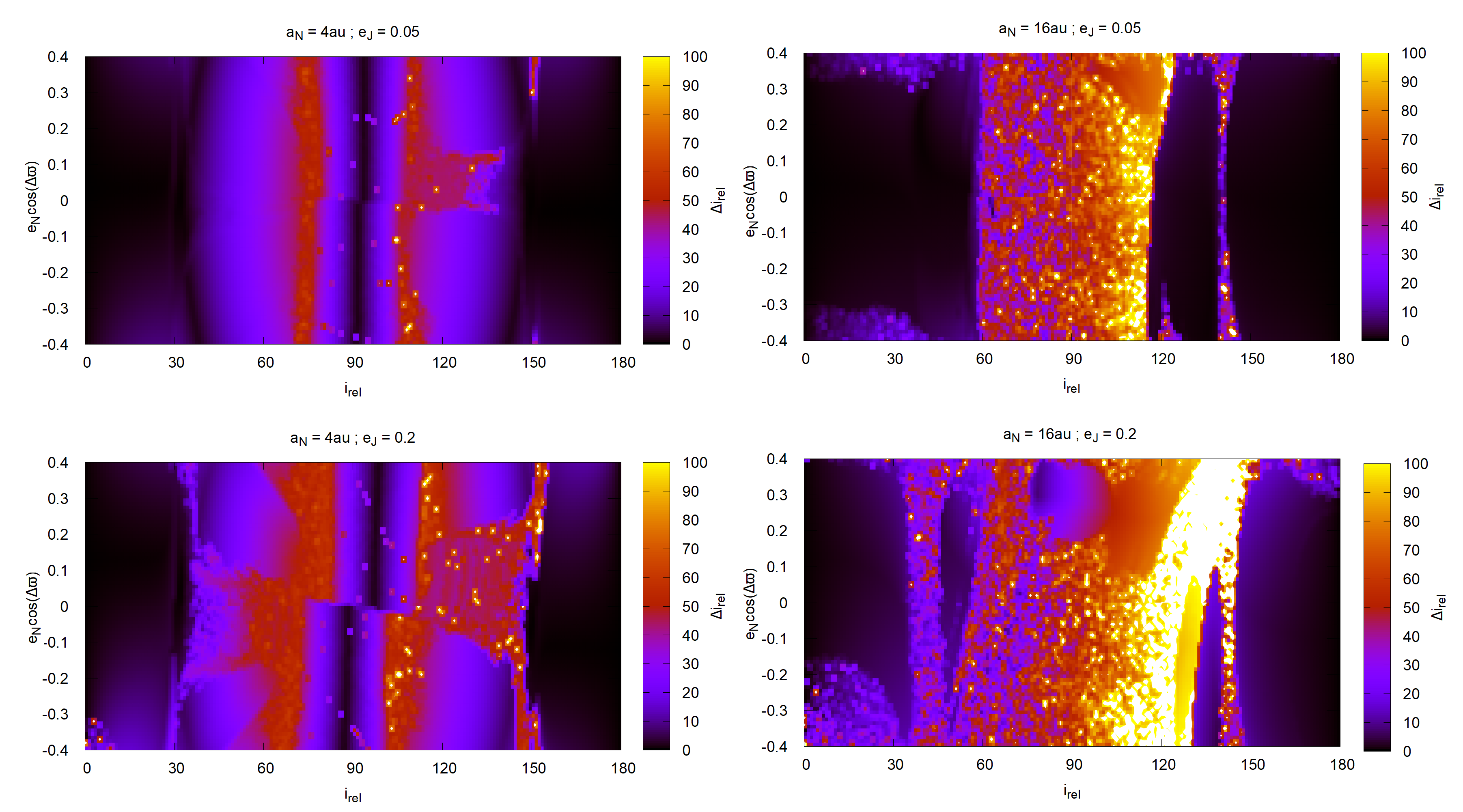}
	\caption{Same as Fig. \ref{deltae} showing changes $\Delta i$.}
	\label{deltai}
\end{figure}

Figs. \ref{deltae} and  \ref{deltai} are dynamical maps that explore the full range of orbital inclinations. They were constructed in similar way than Fig. \ref{dmapae} integrating numerically the exact equations of motions in a time interval of $10^5$ orbital revolutions of our Neptune and calculating maximum changes of $\Delta e_N$ and $\Delta i$ respectively in the grid of  initial  elements ($i,e_N\cos\Delta\varpi$) for $a_N=4$ au  and $a_N=16$ au at left and right columns respectively   and for $e_J=0.05$ and $e_J=0.2$ for up and bottom respectively. 
In all cases it is possible to distinguish the discontinuity around $i \sim 30\degree$ corresponding to the location of the secular resonance. For $i<30\degree$ the system evolves with small changes in eccentricity (Fig. \ref{deltae}) and inclination (Fig. \ref{deltai}) while
for larger inclinations up to $i\sim 150\degree$ the dynamics is more complex and, in the case $a_N=16$ au, unstable with large orbital changes. The dynamics for $i>150\degree$ seems to be analogue to the region $i<30\degree$. These figures represent an expanded exploration of two vertical lines at $a_N=4$ au and  $a_N=16$ au in the upper panel (for the case $\Delta\varpi=0\degree$) and for the lower panel  (for the case $\Delta\varpi=180\degree$) of Fig. \ref{megno} because
the chaotic and regular points of this figure can be directly related with Figs. \ref{deltae} and \ref{deltai}.
From these figures we can also notice that the dynamics for an  exterior  Neptune is completely different than for the case of an  interior  Neptune. Note that for the  interior  Neptune, the quasi polar orbits remain unchanged while for the exterior one large $\Delta i$ are observed. Again we can see that when $\Delta \varpi=0\degree$ (upper part of the plots) there is more stability than $\Delta \varpi=180\degree$ (lower part), specially for the more eccentric Jupiter case.

\section{Conclusions}

We have focused our study on pairs of giant planets of comparable masses that do not fit into the category of hierarchical systems, nor into the category of quasi-circular or quasi-coplanar systems. 
We showed that an eccentric Jupiter planet generates a chaotic region in its neighborhood due to overlap of several MMRs, especially when $\Delta\varpi=180\degree$, which is extended considerably in semi-major axis for $i<30\degree$. Nevertheless, some strong resonances survive  surrounded by chaotic regions, in particular the 1:1.
 Our analysis  of the pseudo equilibrium points for the secular dynamics in  the subspace $(\omega_N,\omega_J)$ indicate that the secular equilibrium configuration $\Delta\varpi=0\degree,180\degree$ typical for low inclination orbits breaks down and change to $\omega_1=\omega_2=\textsc{K} 90\degree$ for some mutual $i<30\degree$ which  depends on $a_1/a_2$ and the eccentricities  (Fig. \ref{adependence}).

 For some mutual critical inclination $i_c$ which in general is between $30\degree$ and $40\degree$ a pericenter secular resonance $g_2=g_1$ takes place dividing in two different dynamical regimes according to the initial mutual inclination.
 For $i<i_c$ the system presents three well defined fundamental frequencies
 that diminish for growing mutual inclination
  and the system evolves with the typical classical secular evolution analogue to the case of low ($e,i$) orbits given by classic theory. In this regime the mutual inclination is almost constant and the
  conservation of the angular momentum is sustained by a regime of forced and free modes that affect the eccentricities and longitudes of pericenter.
 For  $i>i_c$ combinations of the three fundamental frequencies arise in the time evolution of the orbital elements, the mutual inclination changes, the conservation of the angular momentum generates variations in the relative inclination and both eccentricities, following a dynamics of vZLK mechanism imposing large orbital changes that could make the system chaotic and  unstable.
 The ratio between planetary masses does not affect strongly qualitatively the results but has direct effect in the individual fundamental frequencies.
The location of the less massive planet (interior or exterior) with respect to the Jupiter planet generate different dynamics, in particular,
for increasing mutual inclinations the interior Neptune case shows gradual secular changes $\Delta e_N$ and $\Delta i$ while for the  exterior  Neptune case  there is a quick increase in $\Delta e_N$ and $\Delta i$ after the establishment of the vZLK mechanism.  This mechanism occurs at higher inclinations in the case of exterior Neptune, and even at still higher inclinations for the case where $m_N/m_J$ is smaller.
Incidentally, in this sense, the dynamics is reminiscent of the classical vZLK case of a particle perturbed by a planet in circular orbit  where the inclination required for the $\omega$ oscillation is larger when the particle is exterior. 
Note that in all our results (overlap of resonances, chaotic versus regular regions, equilibrium points of $R_s$ and analysis of fundamental frequencies) when the mutual inclination is of the order of 30 degrees there are radical changes in the behavior of the systems, which is in agreement with several works carried out in various theoretical frameworks  including multipole expansions 
\citep{2006Icar..181..555M,2007thesislibert,2011A&A...526A..98F,2019A&A...626A..74V,2023CeMDA.135...22M,bhaskar2024}.  
Finally, as other authors have already pointed out, the secular resonance that occurs for the critical inclination should not be confused with the oscillation of $\Delta\varpi$  that can be observed in quasi-coplanar systems and which is due to the existence of a strong forced mode in the secular evolution.

\textbf{Acknowledgments.} 
We acknowledge funding for the Project  "Dinamica secular y resonante en sistemas planetarios" from CSIC (Udelar). Support from PEDECIBA is also acknowledged.





\bibliographystyle{elsarticle-harv}
\bibliography{refergasu2025}

\end{document}